# Characterisation of the PTB ion counter nanodosimeter's target volume and its equivalent size in terms of liquid H$_2$O

Gerhard Hilgers[1], Thomas Braunroth[1,2], Hans Rabus[1,3]

[1] Physikalisch-Technische Bundesanstalt (PTB), 38116 Braunschweig, Germany
[2] Present Address: Gesellschaft für Anlagen- und Reaktorsicherheit (GRS), 50667 Köln, Germany
[3] Present Address: Physikalisch-Technische Bundesanstalt (PTB), 10587 Berlin, Germany

**Abstract**
The pattern of inelastic interactions in subcellular targets, especially in DNA, is largely responsible for radiation-induced damage to tissue. The track structure of ionising particles is therefore of particular importance in view of the biological effectiveness of ionising radiation. In approaches to model radiation effects to DNA, track structure details on the nanometre scale and the specifics of the target geometry play an important role. Hence, reliable experimental benchmark data for these models require detailed knowledge on the size and shape of the target volume of the devices used to measure quantities related to track structure in simulated nanometric target volumes.

For the first time a dedicated investigation of the target size of a nanodosimeter device has been carried out in order to investigate to what extent measured ionisation cluster size distributions can serve as benchmark data for modelling approaches, with particular focus on the target size in terms of liquid H$_2$O. To this end, measurements with alpha particles from a $^{241}$Am source were carried out using three different target gases, H$_2$O, C$_3$H$_8$ and C$_4$H$_8$O. For each of the three target gases, three different drift-time windows were applied to realise three different target sizes.

A method has been developed to determine the dimensions of the simulated nanometric target volume in liquid H$_2$O for cylindrical and spherical shape, as often used in approaches to model radiation effects to DNA. Simulations with nanometric targets of dimensions determined with this method agree very well with the corresponding measurements. Scaling of the spatial distribution of the extraction efficiency for different target gases and drift-time windows, which corresponds to the nanodosimeter's target volume, in terms of liquid H$_2$O using ($\rho\lambda_{ion}$)-ratios has also been investigated and proved to yield an estimate of the target volume in liquid H$_2$O.

Thus, it can be concluded that ionisation cluster size distributions measured with a nanodosimeter device are suited as benchmark data for approaches that model radiation induced damage to DNA in nanometric volumes of liquid H$_2$O in simple geometries such as cylinders or spheres, provided that the nanodosimeter's target volume is characterised accordingly. By proper selection of drift-time window length as well as target gas and density a wide range of target volume dimensions in terms of liquid H$_2$O can be realised with the PTB Ion Counter nanodosimeter according to specific requirements of modelling approaches.

## 1. Introduction

The pattern of inelastic interactions in subcellular targets, especially in DNA, is substantially responsible for radiation-induced damage to tissue [1,2]. The track structure of ionising particles is therefore of particular importance in view of the biological effectiveness of ionising radiation [3,4].

Advanced modelling approaches exist for predicting biological outcome based on track structure [5-8]. Such modelling is generally difficult to benchmark, as comparisons are only possible with the final biological outcome, where the modelling involves a large number of intermediate steps that may introduce large uncertainties. The physical part of the simulation chain can, in principle, be benchmarked at least for modelling approaches that aim at linking properties of the spatial pattern of ionising radiation interactions to the final biological outcome. Such modelling approaches generally assume simple geometrical shapes of the target, such as cylinders [3-4,9-15] or spheres [16,17].

Nanodosimeters are radiation detectors measuring the relative frequency distribution of the ionisation cluster size, which is characteristic for the experimentally accessible ionisation component of the particle track structure [5-8]. The ionisation cluster size denotes the number $\nu$ of ionisations created in a target volume by a primary particle and all its secondaries. A primary particle can either traverse the target volume or pass it at a distance $d$ with respect to its centre as shown in Fig. 1 (Fig. 1 from [18]). The ionisation cluster size produced in the target is a stochastic quantity that results from the superposition of the pattern of ionisations within a particle track and the geometrical characteristics of the target volume. The ionisation cluster size distribution (ICSD) represents the statistical distribution of the probabilities $P_\nu(Q,d)$ that exactly $\nu$ ions are created in the target volume for radiation quality $Q$ with the primary particles passing the target volume at a distance $d$. Often the mean ionisation cluster size $M_1(Q,d)$ is of particular interest, which is defined by:

$$M_1(Q,d) = \sum_{\nu=0}^{\infty} \nu \cdot P_\nu(Q,d) \ . \qquad (1)$$

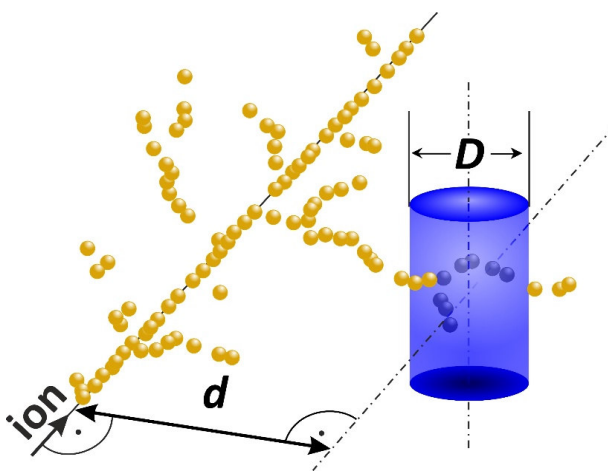

**Figure 1**. Schematic representation of the creation of an ionisation cluster by an ionising particle passing by a cylindrical target volume of diameter $D$ at a distance $d$ from the cylinder axis. In the segment of the particle track shown, the solid circles represent the locations of ionisation interactions.

The ICSD depends, on the one hand, on the specifics of the primary particle (i.e. the radiation quality $Q$) and, on the other hand, on the characteristics of the target. The latter category splits further into characteristics of the target material (target gas and density) and characteristics of the target geometry (shape and size). In approaches to model radiation effects to DNA, specifics of the target geometry play an important role [9,10,17]. Therefore, detailed knowledge on the target size of devices, which are capable of measuring ICSDs in simulated nanometric target volumes, is required to provide reliable benchmark data for modelling radiation effects to DNA.

The measurements in this work were carried out using the PTB ion counter nanodosimeter with three different target gases, $H_2O$ (water vapour), $C_3H_8$ (propane) and $C_4H_8O$ (tetrahydrofuran). The arguments leading to this selection of gases are discussed in [19]. For each of the three target gases, three different target sizes were realised by means of applying three different drift-time windows.

For characterisation of the measured ICSDs with respect to simple target geometries, such as cylindrical or spherical targets, simulations were carried out using the respective geometrical shapes. The parameters determining the cylinder volume, $D_{eff}$ and $H_{eff}$, or the sphere volume, $D_{eff}$, were scanned over a range of values. By minimisation of two different metrics for the deviation between measured and simulated ICSDs, those parameter values that gave the best agreement between measurement and simulation were determined.

## 2. Materials and Methods

### 2.1 PTB ion counter: setup of the experiment

The original setup of the experiment is described in detail in [20]. Later improvements regarding the data acquisition system and the data evaluation procedure as well as an improved characterisation of the device are described in [21].

The nanodosimeter consists of an interaction region filled with a rarefied target gas, an electrode system to extract target gas ions from the interaction region, an evacuated acceleration stage with an ion-counting detector at its end, and a primary particle detector as shown in Fig. 2 (Fig. 2 from [18]). The interaction region, which is located between the electrodes of a plane parallel-plate capacitor, is filled with the target gas at a pressure in the order of 1 mbar. A primary ion traversing the interaction region between the two electrodes is registered by the primary particle detector.

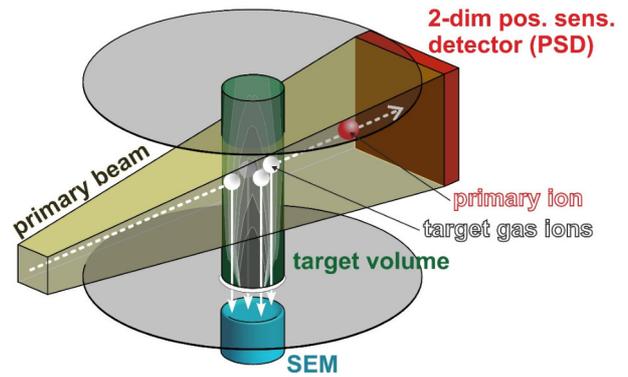

**Figure 2.** Schematic representation of the ion-counting nanodosimeter. An ionising particle passes through the interaction region between the electrodes of a plane parallel-plate capacitor, creating ions in the target gas. These ions drift towards the lower electrode due to an electrical field applied across the electrodes of the capacitor. Ions created within the target volume are extracted via a small aperture in the lower electrode and are detected in a secondary electron multiplier (SEM). The primary ion is registered by the position sensitive detector (PSD) together with its point of impact on the PSD surface. The shading inside the target volume indicates the spatial distribution of the extraction efficiency for the target gas ions created within the target volume.

The ionised target gas molecules generated by the primary particle and its secondaries drift towards the lower electrode due to the electrical field, which is applied across the plane parallel-plate capacitor. Ions passing through the aperture in the lower electrode are extracted from the interaction region. They are then transported through ion optics to an ion counting secondary electron multiplier (SEM), where they are individually detected, and their arrival times recorded.

To enable imaging of the spatial distribution of the extraction efficiency in dependence on drift-time windows, the original trigger detector was replaced by a two-dimensional position sensitive detector (PSD). The active area of the PSD used in the measurements was nominally 20 mm in length and 20 mm in width (First Sensor, DL400-7 [22]). The PSD is not pixel-based, but rather covered with resistive layers on the front and back sides of the silicon chip that are contacted along the "horizontal" and the "vertical" axes, respectively. Position detection in both resistive layers works according to the charge division principle (for details see [18]). Thus, in data processing virtual rectangular pixels of arbitrary size can be configured by choosing independently the dimensions for the two directions. The uncertainties associated with the measurement of the ICSDs and the imaging of the target volume due to the imaging properties of the PSD have been discussed in [18,21].

The alignment of the nanodosimeter was such that primary ions hitting the PSD centrally had a trajectory that was parallel to the electrodes and intersected their rotational-symmetry axis. The coordinates of the primary particle's impact on the PSD surface, together with the known location of the ion source, allows the reconstruction of the primary ion's trajectory and its intersection with a plane that is parallel to the PSD surface and contains the rotational symmetry axis of the electrodes and the extraction aperture. This plane is referred to as the target plane, where the horizontal and vertical coordinates in this plane are designated throughout this paper as $d$ and $h$, respectively. A trajectory ending at the centre of the PSD corresponds to $d = 0$ mm and $h = 0$ mm.

As the investigations in this paper are based on the data sets presented in [19], all relevant details on these data sets can be found there. The data sets consist of records of the arrival positions of the primary ions on the PSD and the drift times of the related secondary ions. Thus, by applying filters on the drift time and primary particle impact position, different target volume sizes and irradiation geometries can be realised in the data analysis. Data obtained in this way are referred to as experimental data in this paper.

## 2.2 PTB ion counter: definition of the target volume

The probability $p(r,z,t_d)$ that a target gas ion passes the extraction aperture after a drift time $t_d$ depends on the position $(r,z)$ where the ion is produced, where $r$ denotes the radial distance from the axis of the extraction aperture and $z$ the distance from the lower electrode. This probability is primarily determined by the electrical field strength as well as by the target gas and its density. The spatial distribution of $p$, which has rotational symmetry around the central axis of the extraction aperture, decreases with increasing distance $z$ from the extraction aperture and with increasing radial distance $r$ from the symmetry axis (Fig. 7 in [18]). The integral of $p(r,z,t_d)$ over the drift time $t_d$ is the "full window" extraction efficiency $\eta_{fw}(r,h)$, where the vertical coordinate $h$ introduced in the previous sections relates to $z$ via $h = z - z_0$, where $z_0$ denotes the height above the lower electrode of a particle trajectory striking the PSD centrally. Since the average drift time $t_d$ depends on the distance $z$ of the target gas ion's point of creation from the extraction aperture (see inset in Fig. 5(a) in [19]), setting a drift-time window $t_c - \Delta t/2 \leq t_d \leq t_c + \Delta t/2$ cuts out a slice from the spatial distribution of $\eta_{fw}(r,z - z_0)$ between $z_{min}$ and $z_{max}$ (corresponding to $t_c - \Delta t/2$ and $t_c + \Delta t/2$, respectively). This slice defines an ion extraction efficiency $\eta(r,h|t_c,\Delta t)$, which depends on the spatial distribution of $\eta_{fw}(r,h)$ and on the width $\Delta t$ and the centre position $t_c$ of the drift-time window. The length of the drift-time window determines the "thickness" $z_{max} - z_{min}$ of the cut-out slice of $\eta_{fw}(r,h)$, i.e. increasing (decreasing) the drift-time window's length increases (decreases) the "thickness" of the slice. Shifting the drift-time window towards larger (smaller) drift time moves the slice towards larger (lower) height above the extraction aperture. Thus, the slice defining the ion extraction efficiency $\eta(r,h|t_c,\Delta t)$ determines the position, shape and size of the target volume.

For each of the three target gases used in this investigation three different drift-time windows were applied, leading to three different target sizes. The centres of the drift-time windows were chosen such that the centres of the target volumes were located at $d = 0$ mm and $h = 0$ mm. A further requirement was that the borders of their projections were sufficiently far away from the PSD's edges to prevent distortions of the images [18]. Due to the different drift times of the three target gases, the parameters of the drift-time windows are different. The parameters of the drift-time windows are listed in table 1. DW-1 denotes the longest and DW-3 the shortest drift-time window. Due to the slightly asymmetric drift time distributions found for all target gases (see Fig. 5(b) in [19]) and the discretisation of the drift time histograms (1 µs for H$_2$O and C$_3$H$_8$, 2 µs for C$_4$H$_8$O) the centres of the drift-time windows differ slightly between target gases.

|  | DW-1 | | DW-2 | | DW-3 | |
|---|---|---|---|---|---|---|
|  | $t_c$ / µs | $\Delta t$ / µs | $t_c$ / µs | $\Delta t$ / µs | $t_c$ / µs | $\Delta t$ / µs |
| H$_2$O | 115 | 15 | 114.5 | 11.5 | 114.5 | 6.5 |
| C$_3$H$_8$ | 95 | 16 | 94 | 11 | 93.5 | 7.5 |
| C$_4$H$_8$O | 247 | 53 | 243 | 29 | 243 | 17 |

**Table 1.** Parameters of the three drift-time windows for the three target gases. DW-1 denotes the longest and DW-3 the shortest drift-time window.

The measured ICSDs for the different targets were obtained by using the respective drift-time window and a virtual pixel of arrival positions within ± 0.5 mm of the targets centre position.

In order to compare measured ICSDs with those obtained from Monte-Carlo simulations, the spatial distribution of $\eta(r,h|t_c,\Delta t)$ is required as input for the simulations and was obtained by ion-transport simulations as described in detail in [21,23].

Validation of the obtained spatial distribution was achieved by comparison of the simulated image of $\eta(r,h|t_c,\Delta t)$ for the corresponding drift time windows in the plane of the PSD surface. The simulation of the image of $\eta(r,h|t_c,\Delta t)$ essentially consists of Monte-Carlo integrations of $p(r,z,t_d)$ along the primary particle's trajectory, which ends on a specified virtual pixel on the PSD surface. To approximate the experimental conditions as close as possible, the simulation of the image on the PSD took into account the influence of (i) the geometrical setup (i.e. position and size of ion source and virtual detecting pixel), (ii) the position resolution of the two-dimensional PSD and (iii) the radial distribution of ionisations due to secondary electrons in the penumbra of the [241]Am alpha particle tracks. However, since the divergence of the primary beam is below 10 mrad, the resulting image is practically identical to the Abel transform of the spatial distribution of $\eta(r,h|t_c,\Delta t)$. Further details can be found in the supplement section.

The effective target height $H_{eff}$ was obtained using Monte-Carlo integration of the simulated extraction efficiency $\eta(r,h|t_c,\Delta t)$ along the symmetry axis, i.e. at $r = 0$ mm. Integration of $\eta(r,h|t_c,\Delta t)$ along the radial direction at $h = 0$ mm gave the effective target diameter $D_{eff}$. The statistical uncertainty of both $H_{eff}$ and $D_{eff}$ was less than 0.72% for all drift time windows, and the statistical uncertainty of their ratio, $F = H_{eff} / D_{eff}$, was below 1% for all drift time windows.

### 2.3 Scaling of the spatial distribution of the extraction efficiency

In order to obtain spatial distributions of the extraction efficiency in terms of liquid $H_2O$, all three linear spatial dimensions were scaled according to the scaling procedure described in [24] using the scaling factors applied in [19]. Strictly speaking, the scaling procedure is only valid for ionisations due to primary alpha particles, however, in close vicinity of the alpha particle trajectory these ionisations are dominating.

In the scaling procedure described in [24], different materials are considered by multiplying by the ratio $(\rho\lambda_{ion})_{H2O} / (\rho\lambda_{ion})_{C3H8}$. Here, $\lambda_{ion}$ is the mean free path for ionisation of the respective materials calculated from the known ionisation cross sections and $\rho$ denotes the density. Based on the revised cross sections for $C_3H_8$ reported in [25], a ratio $(\rho\lambda_{ion})_{H2O} / (\rho\lambda_{ion})_{C3H8}$ = 1.45 is obtained. Together with the densities of liquid $H_2O$ and 1.2 mbar $C_3H_8$ at room temperature of 1 g cm$^{-3}$ and 2.17 µg cm$^{-3}$, respectively, this leads to a scaling factor of $3.15\cdot10^{-6}$ for the dimensions of the spatial distributions $\eta(r,h|t_c,\Delta t)$ simulated for 1.2 mbar $C_3H_8$, such that 1 mm in 1.2 mbar $C_3H_8$ corresponds to 3.15 nm in liquid $H_2O$.

As for $C_4H_8O$, no data of ionisation cross sections for 3.5 MeV alpha particles were available. The scaling factor was therefore estimated using stopping powers from SRIM [26] and mean ionisation energies from the ESTAR data base [27], which leads to a ratio of $(\rho\lambda_{ion})_{C3H8} / (\rho\lambda_{ion})_{C4H8O} \approx 0.73$ [19]. This ratio, together with that of $(\rho\lambda_{ion})_{H2O} / (\rho\lambda_{ion})_{C3H8}$ = 1.45 leads to the ratio $(\rho\lambda_{ion})_{H2O} / (\rho\lambda_{ion})_{C4H8O}$ = 1.06. The densities of liquid $H_2O$ and 1.2 mbar $C_4H_8O$ at room temperature of 1 g cm$^{-3}$ and 3.5 µg cm$^{-3}$, respectively, result in a scaling factor of $3.71\cdot10^{-6}$ for the dimensions of the spatial distributions $\eta(r,h|t_c,\Delta t)$ simulated for 1.2 mbar $C_4H_8O$.

For the scaling between liquid $H_2O$ and $H_2O$ vapour, the same procedure as for $C_4H_8O$ was carried out. From the ASTAR database [27], the mean ionisation energies of liquid $H_2O$ and $H_2O$ vapour are 75 eV and 71.6 eV, respectively. From SRIM [26] the stopping powers of liquid $H_2O$ and $H_2O$ vapour differ by less than 0.5%, leading to $(\rho\lambda_{ion})_{H2O,liquid} / (\rho\lambda_{ion})_{H2O,vapour} \approx 1.05$. With the densities of liquid $H_2O$ and 1.2 mbar $H_2O$ at room temperature of 1 g cm$^{-3}$ and 0.89 µg cm$^{-3}$, respectively, a scaling factor of $0.93\cdot10^{-6}$ is obtained for the dimensions of the spatial distributions $\eta(r,h|t_c,\Delta t)$ simulated for 1.2 mbar $H_2O$ vapour.

### 2.4 Track structure simulations

For comparison with the experimental data, alpha particle tracks in 1.2 mbar $C_3H_8$ gas were simulated using the Monte-Carlo code PTra [5,25] and in liquid water using GEANT4-DNA, version 10.04.p02 [28-31]. The choice of codes was based on the available cross section data implemented in the respective codes.

With both codes, two types of simulation were performed. In the first type, the spatial distribution of the extraction efficiency $\eta(r,h|t_c,\Delta t)$ was used in a Russian-roulette approach to decide whether a secondary ion produced by interactions was detected. In the second approach, the targets were cylinders or spheres for which the probability for counting an ionisation was unity when it occurred inside the cylinder (sphere) and zero if it occurred outside. In both kinds of simulation, the beam of primary alpha particles was traversing the target centrally.

In the PTra $C_3H_8$ track structure code, the scoring of ionisations was implemented in the code and performed during the track simulations. For GEANT4-DNA, the alpha particle tracks were simulated in a slab of liquid $H_2O$ of thickness 270 nm, and all ionisations in a layer between 196 nm and 212 nm were recorded together with their respective coordinates to allow for superposition of particle track structure and target geometry. The simulation results were then processed with a dedicated program, in which the particle tracks were superimposed with the extraction efficiency for target cylinders or spheres to determine the ICSD.

For both simulation codes, the primary particles were alpha particles of 3.5 MeV, which corresponds to the mean energy of the alpha particles emerging from the sealed $^{241}$Am source (with a 10 µm mylar foil) used in the measurements.

The beam's cross section and intensity profile in the simulations were chosen such as to take into account the nanodosimeter's geometrical setup, i.e. the size of the source, the distance between target plane and source as well as the distance between target plane and the surface plane of the PSD. For this geometry, a beam of uniform intensity within a square virtual pixel on the PSD surface becomes a beam of almost square shape in the target plane with a trapezoidal intensity profile along the horizontal and vertical axes. The beam profile resulting from a virtual pixel in the PSD centre of size 1 mm x 1 mm corresponds to the profile P-2 shown in Fig. 3.

In the analysis of the track simulations using GEANT4-DNA, this beam profile was taken into account for both the simulations using the extraction efficiency and the simulations using simple target geometries. In the simulations using the extraction efficiency, the beam size was scaled according the scaling procedure. For simulations using cylindrical targets, the FWHM of the beam with profile P-2 was chosen such that the area $A_{bm}$ of a square of side length FWHM fulfilled the condition $A_{bm} / A_{cyl} = A_{pix} / A_{site}$, in order to preserve the ratio between the cross sections of beam (size of the virtual pixel) and site $A_{pix} / A_{site}$. Here $A_{cyl} = H_{eff} \cdot D_{eff}$ is the cylinder cross section, $A_{pix}$ denotes the virtual pixel area and $A_{site}$ is the effective area of the sites obtained by integration of the Monte-Carlo simulated image of the extraction efficiency $\eta(r,h)$. For spherical targets, the beam size was chosen in the same way.

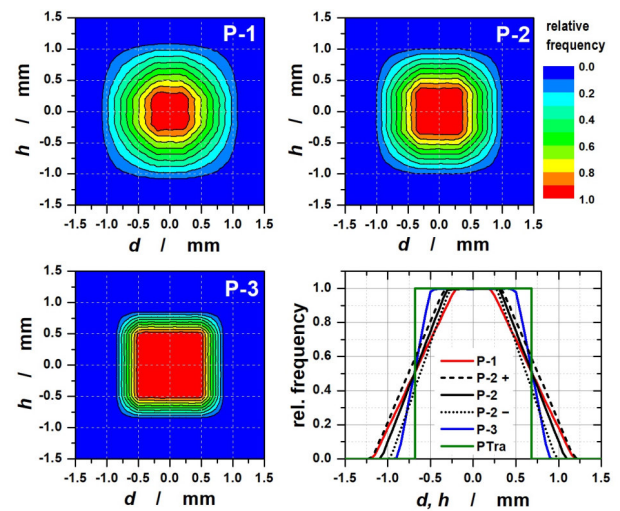

**Figure 3.** Beam profiles applied in the simulations using GEANT4-DNA. Profile P-2 results from a virtual pixel on the PSD of 1 mm x 1 mm. Profiles P-1 and P-3 are of the same FWHM as P-2. Profiles P-2 + and P-2 – base on P-2, but with $d$ and $h$ both scaled to 110% and 90%, respectively. In the simulations using PTra the profile denoted "PTra" was applied.

In the simulations with PTra, the code allowed only the use of a square beam profile; the sides of the square were chosen equal to the FWHM of profile P-2 (profile "PTra" in Fig. 3).

In order to estimate the uncertainty due to the choice of beam profile, simulations with GEANT-DNA were also carried out for two other profiles P-1 and P-3 of the same FWHM as P-2 as well as for two profiles P-2 + and P-2 – that were based on P-2 but with $d$ and $h$ both scaled to 110% and 90%, respectively.

For the estimation of the effect of the ionisation cross sections on the determination of $D_{eff}$ and $H_{eff}$, simulations with GEANT-DNA were carried out using profile P-2 for alpha particles with kinetic energies of 3.25 MeV and 3.75 MeV. At these energies, the stopping powers for alpha particles obtained from SRIM [26] are increased and decreased, respectively, by approximately 5% with respect to the value for 3.5 MeV. Such an approach may serve as an estimate for the uncertainties of the ionisation cross sections.

## 2.5 Optimisation procedure

Scoring of the simulated tracks using simple geometrical shapes was performed for a range of values of the parameters describing the cylindrical volume, $D_{eff}$ and $H_{eff}$, or the spherical volume, $D_{eff}$. The parameter values that gave the best agreement between measurement and simulation were determined by minimisation of two different metrics for the deviation between measured and simulated ICSDs.

The first metric was the reduced $\chi^2$, i.e. $\chi^2(C) / n$, which was calculated according to:

$$\frac{\chi^2(C)}{n} = \frac{1}{n} \sum_{\nu=0}^{\nu_{max}} \frac{\left(P_\nu^{EXP}(C) - P_\nu^{MC}(C)\right)^2}{\left(u\left(P_\nu^{EXP}(C)\right)\right)^2 + \left(u\left(P_\nu^{MC}(C)\right)\right)^2} \quad (2)$$

where $P_\nu^{EXP}(C)$ and $P_\nu^{MC}(C)$ are the measured and simulated data, respectively. $C$ denotes the set of parameters describing the geometry settings, i.e. $t_c$ and $\Delta t$ for the measurements and $H_{eff}$ and $D_{eff}$ for the simulations. The sum includes only those summands, where $P_\nu^{EXP}(C) > 0$ or $P_\nu^{MC}(C) > 0$, and $n$ is the number of data points for which $P_\nu^{EXP}(C) > 0$ or $P_\nu^{MC}(C) > 0$. $u(P_\nu^{EXP}(C))$ and $u(P_\nu^{MC}(C))$ denote the statistical uncertainties of $P_\nu^{EXP}(C)$ and $P_\nu^{MC}(C)$, respectively, which are defined by:

$$u(P_\nu(C)) = \sqrt{\frac{P_\nu(C) \cdot (1 - P_\nu(C))}{N}} \quad (3)$$

$$\text{with:} \quad N = \sum_{\nu=0}^{\nu_{max}} N_\nu(C)$$

$N_\nu(C)$ denotes the absolute frequency of ionisation clusters of cluster size $\nu$.

In the absence of other sources of uncertainties, $\chi^2(C) / n$ would have an expectation value of unity. However, there are uncertainty contributions in the simulations inter alia from the implemented cross section data sets and their processing. Likewise, there are uncertainties in the experiment, e.g. from a residual secondary ion background [18,21]. This background affects mostly the ICSD for large cluster sizes where the probabilities for an ionisation cluster are small. If these additional uncertainties were known quantitatively, they would lead to a reduction of the weights applied to low-frequency cluster sizes. Thus, using only statistical uncertainties may bias the best-fitting target size in favour of infrequent ionisation clusters.

The second metric was the maximum absolute difference $\kappa$ between the two complementary cumulative ICSDs according to

$$\kappa = \max_k \left(\left|F_k^{EXP}(C) - F_k^{MC}(C)\right|\right) \quad (4)$$

which corresponds to the test quantity for a Kolmogorov-Smirnov test of equivalence between (continuous) distributions. $F_k$ is defined by

$$F_k = \sum_{\nu=k}^{\infty} P_\nu \quad (5)$$

In the calculation of both metrics, $\chi^2(C) / n$ and $\kappa(C)$, only those data $P_\nu^{EXP}(C)$ and $P_\nu^{MC}(C)$ were included for which the cumulative probability

$$1 - F_{\nu_c+1}^{EXP}(C) = \sum_{\nu=0}^{\nu_c} P_\nu^{EXP}(C) < 0.99995 \quad (6)$$

$$\text{with:} \quad \nu_c \leq \nu_{max}$$

In principle, the data $P_\nu^{EXP}(C)$ and $P_\nu^{MC}(C)$ for which the cumulative probability is less than 0.00005 should also be excluded. However, in all measured ICSDs $P_0^{EXP}(C)$ exceeds 0.00005.

Both optimisations identify the values of $D_{eff}$ and $H_{eff}$ among those (discrete) values used in the simulations or scoring of simulation results that lead to the best agreement. With the second metric, it is possible to obtain a refined optimisation by linear interpolation of the $F_k(C)$ between adjacent points in the grid of discrete values for $D_{eff}$ and $H_{eff}$ according to

$$F_k(C + \varepsilon \times \Delta C) \approx F_k(C) + \varepsilon \times \frac{F_k(C + \Delta C) - F_k(C)}{\Delta C} \quad (7)$$

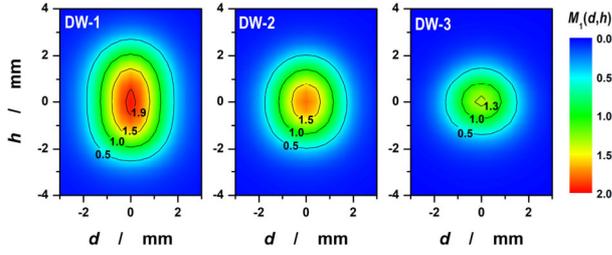

**Figure 4.** Image of the two-dimensional distribution of the mean ionisation cluster size $M_1(d,h)$ as a function of the primary particle trajectory location in the target plane for 1.2 mbar $H_2O$ for the three drift-time windows DW-1, DW-2 and DW-3.

where $\Delta C$ is the step size in the geometry parameters and $\varepsilon$ is a number between 0 and 1. It should be noted that performing the interpolation on $F_k$ rather than $P_\nu$ automatically ensures the normalisation to unity.

## 3. Results and discussion

### 3.1 Influence of the drift-time window on target size and measured ICSDs

Fig. 4 shows the mean ionisation cluster size $M_1$ as a function of the primary particle coordinates in the target plane, $d$ and $h$, for 1.2 mbar $H_2O$ and the three different drift-time windows. Due to the shape of the drift-time distribution (see Fig. 5(b) in [19]), the contour of the target volume is not sharply edged, but rather smoothly sloped. The effect of the drift-time window length on the target size is clearly visible. From the longest drift-time window DW-1 to the shortest window DW-3, there is a noticeable reduction in the extension in $h$ of the target volume.

The maximum of the mean ionisation cluster size $M_1(d,h)$ found in the centre of the target volume at $d = 0$ mm and $h = 0$ mm was also observed to decrease with decreasing length of the drift-time window. As the spatial distribution of $\eta(r,h)$ exhibits rotational symmetry around the central axis of the extraction aperture [18] and as the primary ion beam has negligible divergence, the plots in Fig. 4 approximately represent the Abel-transformed spatial distribution of $\eta(r,h)$ imaged on the target plane, i.e. projections of the (rotationally symmetric) target volume.

In agreement with the change in target size seen in Fig. 4, the measured ICSDs for the three target gases (left: $H_2O$, middle: $C_3H_8$, right: $C_4H_8O$) in Fig. 5 show a reduction in cluster size and mean ionisation cluster size with decreasing length of the drift-time window. ICSDs plotted in red correspond to the longest drift-time window DW-1, the ICSDs plotted in green to the shortest

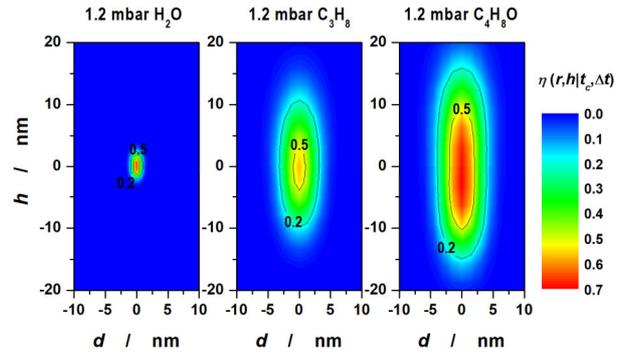

**Figure 6.** Contour plots of central cross-sections of the simulated extraction efficiency $\eta(r,h|t_c,\Delta t)$ for drift-time window DW-1 and 1.2 mbar $H_2O$ (left), 1.2 mbar $C_3H_8$ (middle) and 1.2 mbar $C_4H_8O$ (right). The spatial dimensions are scaled in terms of liquid $H_2O$ using the scaling factors $0.93 \cdot 10^{-6}$ for $H_2O$ vapour, $3.15 \cdot 10^{-6}$ for $C_3H_8$ and $3.71 \cdot 10^{-6}$ for $C_4H_8O$.

window DW-3, and the ICSDs plotted in blue correspond to the medium-length window DW-2.

### 3.2 Scaled spatial distributions of the extraction efficiency

When the scaling procedure is applied to transform the simulated extraction efficiency into the equivalent extraction efficiency in liquid water, pronounced differences between the different target gases are obtained. This is illustrated in Fig. 6, which shows the resulting spatial distributions of the extraction efficiency $\eta(r,h|t_c,\Delta t)$ for the three target gases exemplarily for the longest drift-time window DW-1. The corresponding plots for DW-2 and DW-3 are found in Fig. S2 in the supplement. The dimensions are scaled in terms of liquid $H_2O$ according to the previous discussion. As expected from the scaling factors, the distribution for $H_2O$ vapour is the smallest with diameter $D \approx 3$ nm and height $H \approx 5$ nm, the distribution for $C_4H_8O$ is the largest with diameter $D \approx 12$ nm and height $H \approx 30$ nm and the distribution for $C_3H_8$ is in between with diameter $D \approx 10$ nm and height $H \approx 20$ nm.

### 3.3 Simulations using the spatial distribution $\eta(r,h|t_c,\Delta t)$

#### 3.3.1 Simulations for $C_3H_8$ gas with PTra

Fig. 7 shows the ICSDs measured in 1.2 mbar $C_3H_8$ for the three drift-time windows DW-1, DW-2 and DW-3 as well as those simulated with PTra using the corresponding simulated spatial distributions $\eta(r,h|t_c,\Delta t)$. For all three drift-time windows the measured ICSDs agree reasonably well with the simulations. For all three drift-time windows, the simulated distributions are slightly shifted towards higher cluster sizes. Consequently, the complementary cumulative distributions show differences up to about 7% (see Table 2). The mean ionisation cluster sizes $M_1$ were also found to deviate by as much as 7% between measured and simulated ICSDs.

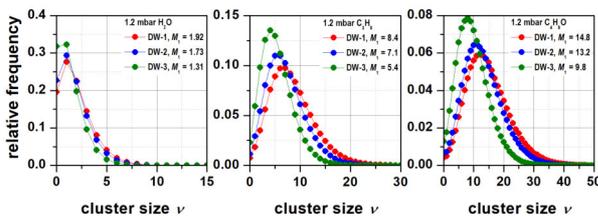

**Figure 5.** ICSDs for the three target gases (left: $H_2O$, middle: $C_3H_8$, right: $C_4H_8O$) and three drift-time windows (red: DW-1, blue: DW-2, green: DW-3).

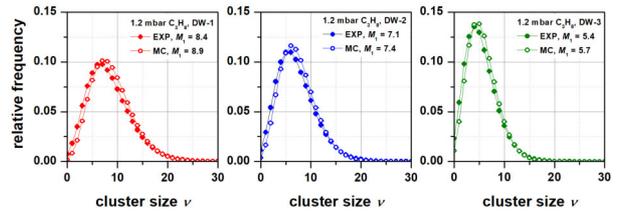

**Figure 7.** ICSDs measured ("EXP") and simulated ("MC") in 1.2 mbar $C_3H_8$ with PTra using the simulated spatial distributions $\eta(r,h|t_c,\Delta t)$ for drift-time windows DW-1 (red, left), DW-2 (blue, middle) and DW-3 (green, right).

### 3.3.2 Simulations for liquid H₂O with GEANT4-DNA

Fig. 8 shows the ICSDs measured in 1.2 mbar H₂O (top row), in 1.2 mbar $C_3H_8$ (middle row) and in 1.2 mbar $C_4H_8O$ (bottom row) for the three drift-time windows DW-1 (red, left column), DW-2 (blue, middle column) and DW-3 (green, right column). The superposition of the track structure in liquid H₂O, which was simulated using GEANT4-DNA, is also shown with the corresponding simulated spatial distributions $\eta(r,h|t_c,\Delta t)$ scaled in terms of liquid H₂O.

Generally, the measured and simulated ICSDs show systematic differences. For all three drift-time windows, the simulated ICSDs for 1.2 mbar H₂O and 1.2 mbar $C_3H_8$ are shifted towards larger cluster sizes compared to the measured ICSDs. This is also observed in the comparison of mean ionisation cluster sizes $M_1$ obtained from the measured and simulated ICSDs, which is reflected in the shift of the peak in the simulated ICSDs towards larger cluster sizes compared to the measurement. For H₂O, the simulated $M_1$ values are 7% to 13% larger than the measured values. For $C_3H_8$, the shift between measured and simulated ICSDs is even larger, where the simulated $M_1$ values are 15% larger than the measured ones. This difference may be an indication that the revision of the cross sections for $C_3H_8$ in [25] may have introduced an overestimation of the ionisation cross sections for alpha particles. For 1.2 mbar $C_4H_8O$, measured and simulated mean ionisation cluster sizes $M_1$ deviate by less than 10%, where the simulated $M_1$ values are smaller than those measured. For large cluster sizes, however, the measured ICSDs show larger frequencies than the simulation in 1.2 mbar $C_4H_8O$ for all three drift-time windows. The deviations between measurement and simulations can be attributed to uncertainties in the scaling factors, which depend on $(\rho\lambda_{ion})$-ratios for the different target gases, as well as any imperfections in modelling the spatial distribution $\eta(r,h|t_c,\Delta t)$. Furthermore, the scaling procedure described in [24] is only valid for the ionisations due to the primary alpha particles, and not for the secondary electrons in the penumbra.

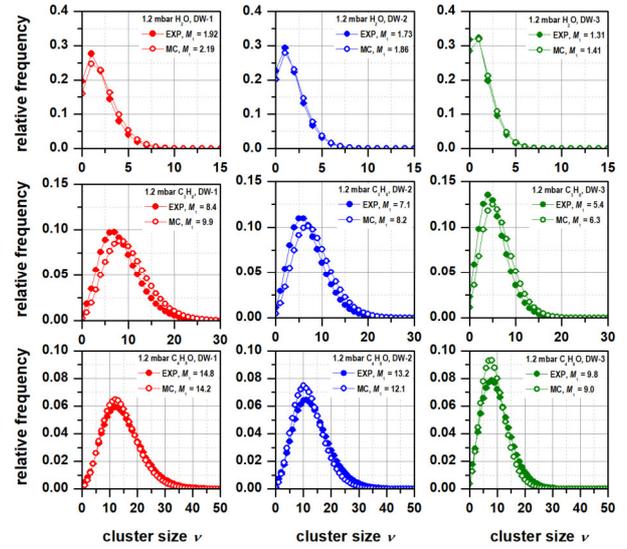

**Figure 8.** ICSDs measured ("EXP") in 1.2 mbar H₂O (top row), in 1.2 mbar $C_3H_8$ (middle row) and in 1.2 mbar $C_4H_8O$ (bottom row) for the three drift-time windows DW-1 (red, left column), DW-2 (blue, middle column) and DW-3 (green, right column) and ("MC") the superposition of the track structure simulated in liquid H₂O with the corresponding simulated spatial distributions $\eta(r,h|t_c,\Delta t)$ scaled in terms of liquid H₂O.

Table 2 shows the values of $\chi^2/n$ and $\kappa$ obtained for the comparison between the measured and simulated ICSDs for the three target gases using GEANT4-DNA and PTra (only for $C_3H_8$) for the three drift-time windows. A general trend for $\chi^2/n$ and $\kappa$ is not observed, neither with respect to the drift-time window width nor with respect to the target gas. Within the simulation data set obtained with GEANT4-DNA both $\chi^2/n$ and $\kappa$ are largest for $C_3H_8$. From PTra simulations for $C_3H_8$, however, both $\chi^2/n$ and $\kappa$ are smaller by a factor of about 2.5 than those obtained for simulations with GEANT4-DNA, indicating a better agreement between measurement and simulation for the simulations using PTra. The same is observed for the mean ionisation cluster sizes $M_1$: $M_1$ values from the simulations with PTra agree significantly better with measured $M_1$ than the $M_1$ from simulations with GEANT4-DNA. Both $\chi^2/n$ and $\kappa$ for simulations using PTra for $C_3H_8$ fall in the same range as those obtained in simulations with GEANT4-DNA for H₂O and $C_4H_8O$.

The large numerical values of $\chi^2/n$ are due to the inclusion of only the statistical uncertainties of $P_\nu^{EXP}(C)$ and $P_\nu^{MC}(C)$ in eq. (2). The sum of the squares of the uncertainties builds the denominator in eq. (2), which is numerically much smaller than the numerator, thus increasing the numerical value of each summand of the sum in eq. (2). Further details can be found in the supplement (Fig. S8). $\kappa$ may therefore better serve as a quantitative measure of the deviation between the two complementary cumulative ICSDs.

|  | DW-1 | | DW-2 | | DW-3 | |
|---|---|---|---|---|---|---|
|  | $\chi^2/n$ | $\kappa$ | $\chi^2/n$ | $\kappa$ | $\chi^2/n$ | $\kappa$ |
| H₂O (G4-DNA) | 782 | 0.066 | 277 | 0.039 | 342 | 0.037 |
| $C_3H_8$ (G4-DNA) | 1101 | 0.139 | 961 | 0.120 | 1086 | 0.114 |
| $C_3H_8$ (PTra) | 367 | 0.068 | 341 | 0.057 | 449 | 0.058 |
| $C_4H_8O$ (G4-DNA) | 102 | 0.036 | 306 | 0.075 | 436 | 0.074 |

**Table 2.** $\chi^2/n$ and $\kappa$ for the comparison between the measured and simulated ICSDs for the three target gases using GEANT4-DNA (G4-DNA) and PTra (only for $C_3H_8$) for the three drift-time windows DW-1, DW-2 and DW-3.

The large values of $\kappa$, i.e. of the maximum deviation between the complementary cumulative distributions, show that simulations using the scaled spatial distributions of the extraction efficiency can reproduce the measured ICSDs only to a limited extent. This may in part be due to the coarse grid used for the ion-transport simulations of 0.4 mm horizontally and 0.84 mm vertically as well as the linear interpolation scheme used for interpolation between grid points. Most likely, the large values of $\kappa$ are due to uncertainties in the scaling factors, which depend on $(\rho\lambda_{ion})$-ratios for the different target gases, as well as imperfections in the modelling of the spatial distribution $\eta(r,h|t_c,\Delta t)$. Moreover, the scaling procedure described in [24] is valid only for the ionisations due to the primary alpha particles, but not for the secondary electrons of the penumbra.

### 3.4 Optimisation for cylindrical targets
### 3.4.1 Simulations for C$_3$H$_8$ gas with PTra

Comparison of ICSDs measured in 1.2 mbar C$_3$H$_8$ for the three drift-time windows and the corresponding ICSDs simulated with PTra using an ideal cylinder of dimensions $D_{eff}$ and $H_{eff}$ as target volume can be seen in Fig. 9, which shows contour plots of the corresponding $\kappa(D_{eff},H_{eff})$ (upper row) and $\chi^2(D_{eff},H_{eff})/n$ (lower row) for DW-1, DW-2 and DW-3. The red circle with the white centre marks $D_{eff}$ and $H_{eff}$ obtained by integrating the spatial distribution of the simulated extraction efficiency $\eta(r,h|t_c,\Delta t)$ for the corresponding drift-time window. Along the red line, the ratio of $D_{eff}$ and $H_{eff}$ is identical to that obtained by integration of $\eta(r,h|t_c,\Delta t)$.

In all three plots, the dependence of $\kappa(D_{eff},H_{eff})$ and $\chi^2(D_{eff},H_{eff})/n$ on $H_{eff}$ is much less pronounced than that on $D_{eff}$, i.e. the simulated ICSDs vary only slightly with varying $H_{eff}$, whereas small variations of $D_{eff}$ lead to significant changes in $\kappa(D_{eff},H_{eff})$ and $\chi^2(D_{eff},H_{eff})/n$. For small $H_{eff}$, as can be seen in the plots for DW-2 and DW-3, there is also a stronger dependence on $H_{eff}$, such that a variation in $D_{eff}$ can be compensated to some extent by an appropriate change in $H_{eff}$. For DW-1 and DW-2, the point of $D_{eff}$ and $H_{eff}$ obtained by integration of $\eta(r,h|t_c,\Delta t)$ lies almost at the bottom of the valleys of $\kappa(D_{eff},H_{eff})$ and $\chi^2(D_{eff},H_{eff})/n$. For DW-3, this point is shifted out of the valley towards smaller values of $D_{eff}$ and $H_{eff}$. Due to the different ratios of $D_{eff}/H_{eff}$ for the three drift-time windows the red line indicating the respective ratio crosses the valleys of $\kappa(D_{eff},H_{eff})$ and $\chi^2(D_{eff},H_{eff})/n$ at different positions $D_{eff}$ and $H_{eff}$.

Fig. 10 shows the Kolmogorov-Smirnov test of equivalence between the measured ICSD for the drift-time window DW-1 and simulated ICSDs with $D_{eff}$ varied along the straight line in the left panels of Fig. 9. The plot on the left shows the direct comparison between measured and simulated ICSDs. From this comparison the best visual agreement is found for the simulation with $D_{eff}$ = 0.95 mm, which almost coincides with the measurement. The right plot shows the difference between the measured complementary cumulative ICSD and the simulated ones with the respective maximum difference indicated by the full circles. The smallest differences are found for the simulation with $D_{eff}$ = 0.95 mm. The inset in the right plot shows $\kappa$, i.e. the absolute maximum difference between the measured and simulated complementary cumulative ICSDs, dependent on $D_{eff}$. At $D_{eff}$ = 0.95 mm, $\kappa$ shows a distinct minimum, indicating best agreement between simulation and measurement.

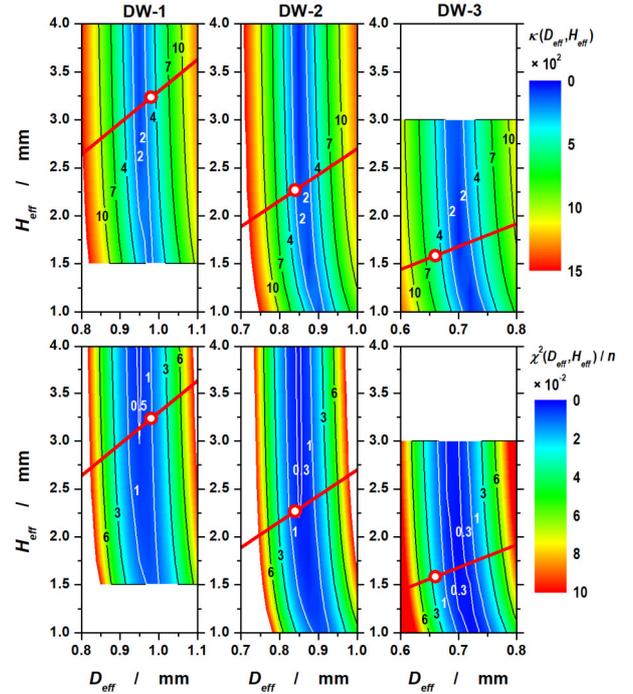

**Figure 9.** Contour plots of $\kappa(D_{eff},H_{eff})$ (upper row) and $\chi^2(D_{eff},H_{eff})/n$ (lower row) for the comparison of ICSDs measured in 1.2 mbar C$_3$H$_8$ for the three drift-time windows and the corresponding ICSDs simulated with PTra using an ideal target cylinder of dimensions $D_{eff}$ and $H_{eff}$. The red circle with the white centre marks $D_{eff}$ and $H_{eff}$ obtained from integration of the spatial distribution of the simulated extraction efficiency $\eta(r,h|t_c,\Delta t)$ for the corresponding drift-time window. Along the red line, the ratio of $D_{eff}/H_{eff}$ is that obtained from integration of $\eta(r,h|t_c,\Delta t)$.

The left plot in Fig. 11 shows $\chi^2(D_{eff},H_{eff})/n$ and $\kappa(D_{eff},H_{eff})$ for the three drift-time windows. Assuming the ratio of $F = H_{eff} / D_{eff}$ as obtained by integration of $\eta(r,h|t_c,\Delta t)$, the values of $D_{eff}$ and $H_{eff}$ are varied according to this ratio as indicated by the red lines in the plots of Fig. 9, i.e. $H_{eff} = F \cdot D_{eff}$. $F$ is a constant factor for each of the drift-time windows DW-1, DW-2 and DW-3, and the value of $F$ is indicated in the legend of the left-hand side panel of Fig. 11. Both $\chi^2(D_{eff},H_{eff}) / n$ and $\kappa(D_{eff},H_{eff})$ show distinct minima at 0.70 mm (DW-3, $F = 2.4$), at 0.85 mm (DW-2, $F = 2.7$) and at 0.95 mm (DW-1, $F = 3.3$). For comparison, the vertical dotted lines indicate the values of $D_{eff}$ obtained by integration of $\eta(r,h|t_c,\Delta t)$. The values of $D_{eff}$ obtained by integration of $\eta(r,h|t_c,\Delta t)$ agree reasonably with the values found at the minima of $\chi^2(D_{eff},H_{eff}) / n$ and $\kappa(D_{eff},H_{eff})$, they differ by less than 10%.

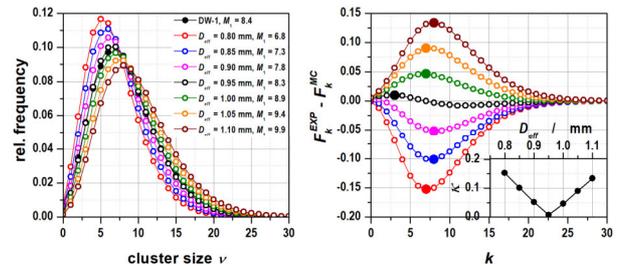

**Figure 10.** Left: Comparison of the experimental ICSD for drift-time window DW-1 and the variation of the simulated ICSDs with parameter $D_{eff}$ along the straight line in the left-hand panel of Fig. 9. Right: Difference between the simulated complementary cumulative ICSDs and the experimental one. The full circles indicate the respective maximum difference. The inset shows $\kappa$ in dependence of $D_{eff}$.

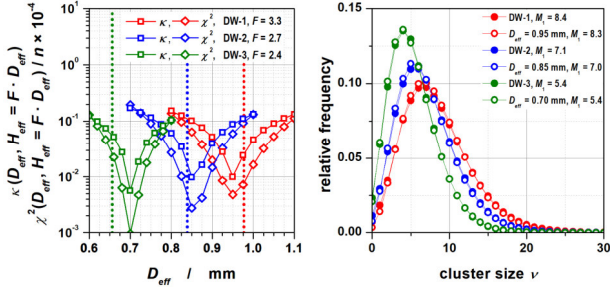

**Figure 11.** Left: $\chi^2(D_{eff},H_{eff}) / n$ and $\kappa(D_{eff},H_{eff})$ for drift-time windows DW-1, DW-2 and DW-3. $D_{eff}$ and $H_{eff}$ are varied according to $H_{eff} = F \cdot D_{eff}$ with the ratio $F = H_{eff} / D_{eff}$ obtained by integration of $\eta(r,h|t_c,\Delta t)$. The vertical dotted lines indicate the values of $D_{eff}$ obtained by integration of $\eta(r,h|t_c,\Delta t)$. Right: ICSDs measured in 1.2 mbar C$_3$H$_8$ for drift-time windows DW-1, DW-2 and DW-3 (full circles) and the corresponding ICSDs simulated with PTra using an ideal cylinder with the values found for $D_{eff}$ and $H_{eff}$ at the minimum of $\chi^2(D_{eff},H_{eff}) / n$ and $\kappa(D_{eff},H_{eff})$ for the respective drift-time window (open circles).

In Fig. 11, the right plot shows the ICSDs measured in 1.2 mbar C$_3$H$_8$ for the three drift-time windows and the corresponding ICSDs simulated with PTra using an ideal cylinder with the values found for $D_{eff}$ and $H_{eff}$ at the minima of $\chi^2(D_{eff},H_{eff}) / n$ and $\kappa(D_{eff},H_{eff})$ for the respective drift-time window, i.e. 0.70 mm for DW-3, 0.85 mm for DW-2 and 0.95 mm for DW-1. The agreement between measured and simulated ICSDs is very good, with maximum difference between measured and simulated complementary cumulative ICSDs $\kappa \lesssim 1\%$. For cluster sizes $\nu \lesssim 20$ measurement and simulation almost coincide, except for larger cluster sizes where deviations between measured and simulated ICSDs occur. The mean ionisation cluster sizes $M_1$ are almost identical.

In the contour plots shown in Fig. 9, both $\chi^2(D_{eff},H_{eff}) / n$ and $\kappa(D_{eff},H_{eff})$ are almost independent on $H_{eff}$ for large values of $H_{eff}$. Furthermore, there is little variation of $\chi^2(D_{eff},H_{eff}) / n$ and $\kappa(D_{eff},H_{eff})$ with both parameters in the bottom of the valley. Therefore, a large variety of values of $D_{eff}$ and $H_{eff}$ leads to small $\chi^2(D_{eff},H_{eff}) / n$ and $\kappa(D_{eff},H_{eff})$, making a unique determination of $D_{eff}$ and $H_{eff}$ impossible. However, the integration of $\eta(r,h|t_c,\Delta t)$ yields additional information on $D_{eff}$ and $H_{eff}$, in particular the ratio $H_{eff} / D_{eff}$. Applying this ratio in the comparison of measured and simulated ICSDs leads to unique results for $D_{eff}$ and $H_{eff}$ which are consistent with the shape of $\chi^2(D_{eff},H_{eff}) / n$ and $\kappa(D_{eff},H_{eff})$. Therefore, the application of the ratio $H_{eff} / D_{eff}$ obtained by integration of the spatial distribution of the extraction efficiency $\eta(r,h|t_c,\Delta t)$ seems justified.

### 3.4.2 Simulations for liquid H$_2$O with GEANT4-DNA

Fig. 12 shows, exemplarily for the drift-time window DW-2, contour plots of $\kappa(D_{eff},H_{eff})$ resulting from the comparison of ICSDs measured in 1.2 mbar H$_2$O (left), in 1.2 mbar C$_3$H$_8$ (middle) and in 1.2 mbar C$_4$H$_8$O (right) as well as the corresponding ICSDs simulated with GEANT4-DNA using an ideal cylinder of liquid H$_2$O with dimensions $D_{eff}$ and $H_{eff}$ as target volume. The contour plots of $\chi^2(D_{eff},H_{eff}) / n$ are omitted here and can be found in the supplement (Fig. S4). As in Fig. 9, the red circle with the white centre marks $D_{eff}$ and $H_{eff}$, now scaled in terms of liquid H$_2$O, that were obtained by integration of $\eta(r,h|t_c,\Delta t)$ for drift-time window DW-2. Along the red line, the ratio of $D_{eff}$ and $H_{eff}$ is the same as the one obtained by integration of $\eta(r,h|t_c,\Delta t)$. The scaling factors applied to scale the dimensions of $D_{eff}$ and $H_{eff}$ correspond to those discussed previously.

Similar to the plots shown in Fig. 9, for all target gases the dependence of $\kappa(D_{eff},H_{eff})$ on $H_{eff}$ is much less pronounced than on $D_{eff}$. The simulated ICSDs vary only slightly with varying $H_{eff}$, whereas small variations of $D_{eff}$ lead to pronounced changes of $\kappa(D_{eff},H_{eff})$. However, for small $H_{eff}$ a variation in $D_{eff}$ can be compensated to some extent by an appropriate change in $H_{eff}$, as can be seen for 1.2 mbar C$_3$H$_8$ and for 1.2 mbar C$_4$H$_8$O. For 1.2 mbar C$_3$H$_8$, the point of $D_{eff}$ and $H_{eff}$ obtained from integration of $\eta(r,h|t_c,\Delta t)$ is almost at the bottom of the valleys of $\kappa(D_{eff},H_{eff})$. For 1.2 mbar H$_2$O and 1.2 mbar C$_4$H$_8$O, this point is shifted out of the valleys towards larger and smaller values of $D_{eff}$ and $H_{eff}$, respectively.

The plots in the left column of Fig. 13 show for 1.2 mbar H$_2$O (top row), for 1.2 mbar C$_3$H$_8$ (middle row) and for 1.2 mbar C$_4$H$_8$O (bottom row) the $\kappa(D_{eff},H_{eff})$ values for drift-time windows DW-1, DW-2 and DW-3, assuming the ratio of $F = H_{eff} / D_{eff}$ obtained by integration of $\eta(r,h|t_c,\Delta t)$. The plots of $\chi^2(D_{eff}, H_{eff})/n$ are omitted here and can be found in the supplement (Fig. S5). The values of $D_{eff}$ and $H_{eff}$ are varied according to $H_{eff} = F \cdot D_{eff}$ as indicated by the red lines in the plots of Fig. 12. $F$ is constant for each combination of drift-time window and target gas. The value of $F$ is indicated in the legends of the plots. For comparison, the vertical dotted lines indicate the values of $D_{eff}$ obtained by integration of $\eta(r,h|t_c,\Delta t)$ scaled in terms of liquid H$_2$O. The plots in the right column of Fig. 13 show the ICSDs measured in each of the respective target gases for the three drift-time windows and the corresponding ICSDs simulated with GEANT4-DNA using an ideal cylinder of liquid H$_2$O with the values for $D_{eff}$ and $H_{eff}$ obtained at the minimum of $\kappa(D_{eff}, H_{eff})$ for the respective drift-time window.

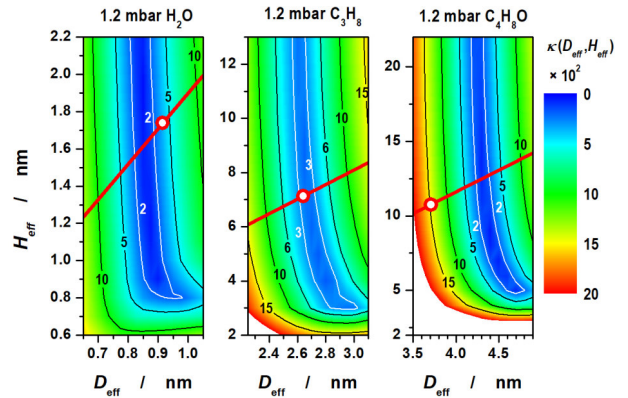

**Figure 12.** Contour plots of $\kappa(D_{eff},H_{eff})$ of the comparison of ICSDs measured in 1.2 mbar H$_2$O (left), in 1.2 mbar C$_3$H$_8$ (middle) and in 1.2 mbar C$_4$H$_8$O (right) for drift-time window DW-2 and the corresponding ICSDs simulated with GEANT4-DNA using an ideal target cylinder of liquid H$_2$O with dimensions $D_{eff}$ and $H_{eff}$. The red circle with the white centre marks $D_{eff}$ and $H_{eff}$, scaled in terms of liquid H$_2$O, obtained by integration of $\eta(r,h|t_c,\Delta t)$ for drift-time window DW-2. Along the red line the ratio of $D_{eff} / H_{eff}$ is the same as it is obtained by integration of $\eta(r,h|t_c,\Delta t)$.

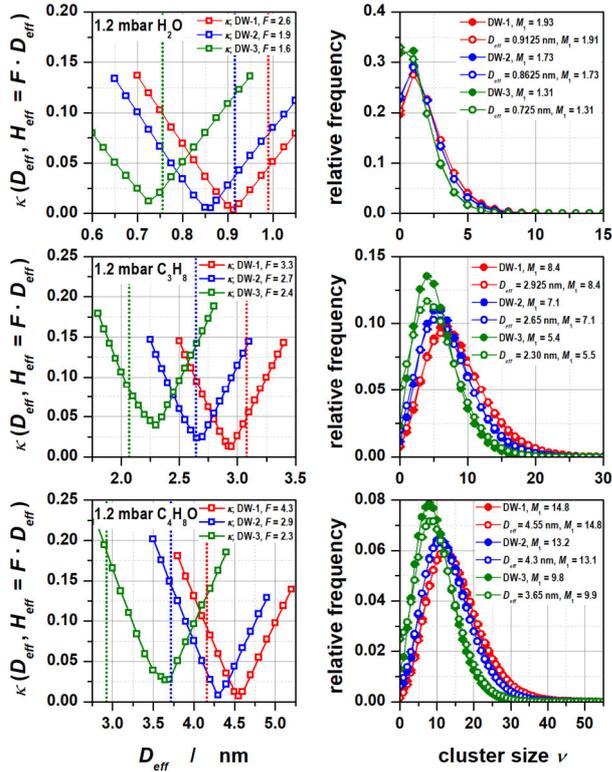

**Figure 13.** $\kappa(D_{eff}, H_{eff})$ for 1.2 mbar H$_2$O (top row), for 1.2 mbar C$_3$H$_8$ (middle row) and for 1.2 mbar C$_4$H$_8$O (bottom row) for drift-time windows DW-1, DW-2 and DW-3. $D_{eff}$ and $H_{eff}$ are jointly varied according to $H_{eff} = F \cdot D_{eff}$ with the ratio $F = H_{eff} / D_{eff}$ obtained from integration of $\eta(r,h|t_c,\Delta t)$. The vertical dotted lines indicate the values of $D_{eff}$ obtained from integration of $\eta(r,h|t_c,\Delta t)$ scaled to liquid H$_2$O. Right: ICSDs measured in the respective target gases for drift-time windows DW-1, DW-2 and DW-3 (full circles) and the corresponding ICSDs simulated with GEANT4-DNA using an ideal cylinder of liquid H$_2$O with the values for $D_{eff}$ and $H_{eff}$ corresponding to the minimum of $\kappa(D_{eff}, H_{eff})$ for the respective drift-time window (open circles).

The plots of $\kappa(D_{eff}, H_{eff})$ for the three target gases and the three drift-time windows show distinct minima for all combinations of target gas and drift-time window. The values of $D_{eff}$ and $H_{eff}$, for which the minima of $\chi^2(D_{eff}, H_{eff})/n$ and $\kappa(D_{eff}, H_{eff})$ are obtained, are listed in the first six columns of table 3. For 1.2 mbar H$_2$O, the minima of $\chi^2(D_{eff}, H_{eff})/n$ and $\kappa(D_{eff}, H_{eff})$ coincide for the three drift time windows, whereas for both 1.2 mbar C$_3$H$_8$ and 1.2 mbar C$_4$H$_8$O, the minima coincide only for drift time windows DW-1 and DW-2. For DW-3 slight differences of less than 2.3% can be seen for $D_{eff}$ and $H_{eff}$ at the position of minima. However, the difference in $D_{eff}$ and $H_{eff}$ corresponds to the increment in the discrete variation of $D_{eff}$ in the simulation. For 1.2 mbar H$_2$O, the values of $D_{eff}$ obtained by integration of the scaled $\eta(r,h|t_c,\Delta t)$ are systematically shifted towards larger values compared to the values found in the minima of $\kappa(D_{eff}, H_{eff})$, whereas for 1.2 mbar C$_4$H$_8$O the shift is systematically in the opposite direction. This indicates that the estimation of the $(\rho\lambda_{ion})$-ratios based on stopping powers and mean ionisation energies is indeed an approximation, however, the values of $D_{eff}$ obtained by integration of the scaled $\eta(r,h|t_c,\Delta t)$ deviate between 5% and 25% from those found at the minima of $\kappa(D_{eff}, H_{eff})$. For 1.2 mbar C$_3$H$_8$, no systematic shift is observed between the values of $D_{eff}$ obtained by integrating the scaled $\eta(r,h|t_c,\Delta t)$ and those found at the minima of $\kappa(D_{eff}, H_{eff})$, they agree within 10%.

A very good agreement between measured and simulated ICSDs is obtained for all target gases. The deviation of the cumulative complementary ICSDs is mostly below 1.5% and as large as 4.0% in the worst case. Measurement and simulation are almost coincident over the whole range of cluster sizes. The mean ionisation cluster sizes $M_1$ agree within 2% for all combinations of target gas and drift-time window.

For the determination of the dimensions $D_{eff}$ and $H_{eff}$ of the target cylinder of liquid H$_2$O, which superimposes the track structure simulated with GEANT4-DNA in liquid H$_2$O, the same discussion applies as in the previous situation, when simulating an ideal target cylinder with PTra. The slight variation of the values of $\chi^2(D_{eff}, H_{eff})/n$ and $\kappa(D_{eff}, H_{eff})$ along the bottom of the valleys makes the unique determination of $D_{eff}$ and $H_{eff}$ impossible. As before, applying the ratio $H_{eff}/D_{eff}$ obtained from integration of $\eta(r,h|t_c,\Delta t)$ in the comparison of measured and simulated ICSDs leads to unique results for $D_{eff}$ and $H_{eff}$ which are consistent with the shape of $\chi^2(D_{eff}, H_{eff})/n$ and $\kappa(D_{eff}, H_{eff})$. Therefore, the application of the ratio $H_{eff}/D_{eff}$ obtained by integration of $\eta(r,h|t_c,\Delta t)$ in this case also seems justified.

### 3.5 Simulation with GEANT4-DNA using spherical target volumes

Some approaches to model radiation effects to DNA apply a simple target geometry using spheres of liquid H$_2$O as target volumes [17]. For the shortest drift-time window DW-3, the Abel-projected spatial distributions of the extraction efficiency show shapes which are close to a circle. For H$_2$O (see Fig. 4), the shape of the target volume is almost spherical, but for C$_3$H$_8$ and C$_4$H$_8$O deviations from the spherical shape are visible (see Fig. S2 in the supplement). To investigate to which extent the target volume of the measured ICSDs can be approximated by a sphere of liquid H$_2$O, the measured ICSDs for the three target gases with DW-3 were compared to ICSDs obtained from track structures, which were simulated with GEANT4-DNA in liquid H$_2$O and superimposed with spherical target volumes of liquid H$_2$O with diameter $D_{eff}$.

| | | DW-1, cylinder | | DW-2, cylinder | | DW-3, cylinder | | DW-3, sphere |
|---|---|---|---|---|---|---|---|---|
| | | $D_{eff}$ / nm | $H_{eff}$ / nm | $D_{eff}$ / nm | $H_{eff}$ / nm | $D_{eff}$ / nm | $H_{eff}$ / nm | $D_{eff}$ / nm |
| H$_2$O | $\kappa$ | 0.9125 | 2.3725 | 0.8625 | 1.6388 | 0.725 | 1.16 | 0.80 |
| | $\chi^2/n$ | | | | | | | |
| C$_3$H$_8$ | $\kappa$ | 2.95 | 9.735 | 2.65 | 7.155 | 2.30 | 5.52 | 2.45 |
| | $\chi^2/n$ | | | | | 2.25 | 5.40 | |
| C$_4$H$_8$O | $\kappa$ | 4.55 | 19.57 | 4.30 | 12.47 | 3.65 | 8.40 | 3.95 |
| | $\chi^2/n$ | | | | | 3.60 | 8.28 | 4.00 |

**Table 3.** Values of $D_{eff}$ and $H_{eff}$ for a cylindrical target volume (first six columns), for which the minima of $\chi^2(D_{eff},H_{eff}) / n$ and $\kappa(D_{eff},H_{eff})$ are obtained for the three target gases and drift-time windows DW-1, DW-2 and DW-3, and values of $D_{eff}$ for a spherical target volume (last column), for which the minima of $\chi^2(D_{eff}) / n$ and $\kappa(D_{eff})$ are obtained for the three target gases and drift-time window DW-3.

The left plot in Fig. 14 shows for 1.2 mbar H$_2$O, for 1.2 mbar C$_3$H$_8$ and for 1.2 mbar C$_4$H$_8$O the $\kappa(D_{eff})$ values for drift-time window DW-3. The plots of $\chi^2(D_{eff},H_{eff}) / n$ are omitted here and can be found in the supplement (Fig. S6). The plot of $\kappa(D_{eff})$ for the three target gases and drift-time window DW-3 shows distinct minima.

The values of $D_{eff}$ for DW-3, for which the minima of $\chi^2(D_{eff}) / n$ and $\kappa(D_{eff})$ are obtained, are listed in the last column of table 3. For 1.2 mbar H$_2$O and 1.2 mbar C$_3$H$_8$, the minima of $\chi^2(D_{eff}) / n$ and $\kappa(D_{eff})$ coincide, whereas for 1.2 mbar C$_4$H$_8$O deviations of less than 1.3% can be seen for $D_{eff}$ at the minima. The difference in $D_{eff}$, however, corresponds to the increment in the discrete variation of $D_{eff}$ in the simulation. For spherical target volumes of all three target gases, $D_{eff}$ is systematically larger by about 9% to 11% than the value obtained for cylindrical target volumes with DW-3. Due to the different shapes, the scored number of ionisations created by secondary electrons in the penumbra of the primary alpha particle tracks is different inside spherical and cylindrical target volumes. Thus, the increased $D_{eff}$ for spherical compared to cylindrical target volumes compensates for the loss of scored ionisations by secondary electrons inside spherical target volumes compared to cylindrical volumes.

The plot on the right-hand side of Fig. 14 shows the ICSDs measured in the respective target gases for drift-time window DW-3 as well as the corresponding ICSDs simulated with GEANT4-DNA using an ideal sphere of liquid H$_2$O, where the values for $D_{eff}$ were those at the minimum of $\kappa(D_{eff})$.

A very good agreement between measured and simulated ICSDs is found for all target gases, and most deviations of the cumulative complementary ICSDs are less than 1%, rising to about 2.5% in the worst case. Measurement and simulation are almost coincident over the whole range of relevant cluster sizes. The mean ionisation cluster sizes $M_1$ also agree within 1% for all target gases.

### 3.6 Estimation of uncertainties

Table 4 shows $D_{eff}$ (bold font) found at the minima of $\chi^2(D_{eff},H_{eff}) / n$ and $\kappa(D_{eff},H_{eff})$ from the comparison of measured ICSDs (in 1.2 mbar H$_2$O, in 1.2 mbar C$_3$H$_8$ and in 1.2 mbar C$_4$H$_8$O for the three drift time windows) and the corresponding simulated ICSDs (obtained with GEANT4-DNA using an ideal cylinder of liquid H$_2$O with dimensions $D_{eff}$ and $H_{eff} = F \cdot D_{eff}$ as target volume). Missing entries in the row denoted with $\chi^2 / n$ indicate identical values for $D_{eff}$ at the minima of $\chi^2(D_{eff},H_{eff}) / n$ and $\kappa(D_{eff},H_{eff})$. The numbers printed in normal font are the values of $\kappa(D_{eff},H_{eff})$ at the corresponding minimum. In the simulations using GEANT4-DNA alpha particles with a mean kinetic energy of 3.5 MeV (to simulate an $^{241}$Am-source sealed with a 10 µm mylar foil) were used for beam profiles P-1, P-2 +, P-2, P-2 − and P-3 and alpha particles of 3.25 MeV and 3.75 MeV kinetic energy were used for beams with profile P-2.

From Table 4, simulations obtained with beam profiles P-1 and P-2 + (P-3 and P-2 −), which are wider (narrower) than the reference profile P-2, have a systematic tendency to produce larger (smaller) values of $D_{eff}$. Due to the larger (smaller) number of particle tracks passing the target cylinder at a large distance from its centre and the smaller (larger) number of particle tracks passing the target cylinder close to its centre for wider (narrower) beam profiles, the mean track length inside the cylinder volume decreases (increases), thus leading to smaller (larger) ionisation cluster sizes. In consequence, this decrease (increase) of large ionisation clusters in the simulated ICSDs is compensated by an increase (decrease) of the target cylinder diameter $D_{eff}$ for best agreement between measurement and simulation. This result is consistent with the finding for the variation of the cross sections for ionisation: above the Bragg peak energy, an increasing (decreasing) alpha particle energy leads to a decreasing (increasing) ionisation cross section, resulting in smaller (larger) ionisation cluster sizes and, consequently, leading to a larger (smaller) target cylinder diameter $D_{eff}$.

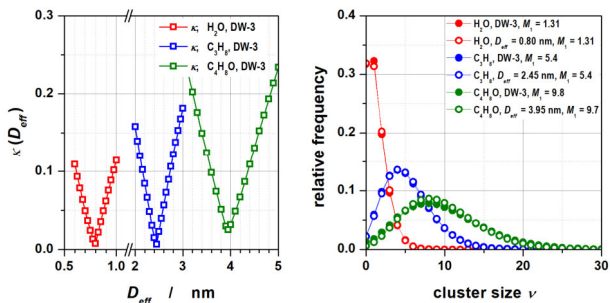

**Figure 14.** Left: $\kappa(D_{eff})$ for 1.2 mbar H$_2$O (red), for 1.2 mbar C$_3$H$_8$ (blue) and for 1.2 mbar C$_4$H$_8$O (green) for drift-time window DW-3. Right: ICSDs measured in the respective target gases for drift-time window DW-3 (full circles) and the corresponding ICSDs simulated with GEANT4-DNA using an ideal sphere of liquid H$_2$O with the values for $D_{eff}$ obtained from the minimum of $\kappa(D_{eff})$ (open circles).

|  |  | H₂O, $D_{eff}$ / nm | | | C₃H₈, $D_{eff}$ / nm | | | C₄H₈O, $D_{eff}$ / nm | | |
|---|---|---|---|---|---|---|---|---|---|---|
|  |  | DW-1 | DW-2 | DW-3 | DW-1 | DW-2 | DW-3 | DW-1 | DW-2 | DW-3 |
| P-1, ²⁴¹Am | $\kappa$ | **0.950** | **0.875** | **0.750** | **3.00** | **2.75** | **2.40** | **4.6** | **4.4** | **3.8** |
|  |  | 0.0128 | 0.0128 | 0.0200 | 0.0194 | 0.0311 | 0.0539 | 0.0068 | 0.0225 | 0.0431 |
|  | $\chi^2/n$ | **0.925** |  |  | **2.95** | **2.70** | **2.25** |  |  | **3.7** |
| P-2 +, ²⁴¹Am | $\kappa$ | **0.950** | **0.900** | **0.775** | **3.05** | **2.80** | **2.40** | **4.7** | **4.5** | **3.9** |
|  |  | 0.0102 | 0.0116 | 0.0173 | 0.0251 | 0.0366 | 0.0596 | 0.0128 | 0.0253 | 0.0516 |
|  | $\chi^2/n$ |  |  | **0.750** | **0.300** | **2.70** | **2.30** | **4.6** | **4.4** | **3.7** |
| P-2, ²⁴¹Am | $\kappa$ | **0.900** | **0.850** | **0.725** | **2.95** | **2.65** | **2.30** | **4.6** | **4.3** | **3.7** |
|  |  | 0.0076 | 0.0057 | 0.0119 | 0.0127 | 0.0231 | 0.0392 | 0.0121 | 0.0078 | 0.0272 |
|  | $\chi^2/n$ | **0.925** |  |  |  |  | **2.25** | **4.5** |  | **3.6** |
| P-2 −, ²⁴¹Am | $\kappa$ | **0.900** | **0.825** | **0.700** | **2.85** | **2.60** | **2.20** | **4.4** | **4.2** | **3.5** |
|  |  | 0.0099 | 0.0074 | 0.0081 | 0.0109 | 0.0104 | 0.0205 | 0.0187 | 0.0102 | 0.0048 |
|  | $\chi^2/n$ |  |  |  |  |  |  | **4.5** |  |  |
| P-3, ²⁴¹Am | $\kappa$ | **0.900** | **0.825** | **0.675** | **2.85** | **2.60** | **2.20** | **4.4** | **4.2** | **3.5** |
|  |  | 0.0105 | 0.0079 | 0.0095 | 0.0094 | 0.0088 | 0.0174 | 0.0205 | 0.0146 | 0.0064 |
|  | $\chi^2/n$ |  |  | **0.700** |  |  | **2.15** | **4.5** |  |  |
| P-2, 3.75 MeV | $\kappa$ | **0.950** | **0.900** | **0.775** | **3.10** | **2.80** | **2.40** | **4.8** | **4.5** | **3.8** |
|  |  | 0.0061 | 0.0033 | 0.0137 | 0.0163 | 0.0206 | 0.0401 | 0.0066 | 0.0111 | 0.0290 |
|  | $\chi^2/n$ |  |  | **0.750** |  | **2.75** | **2.35** |  |  |  |
| P-2, 3.25 MeV | $\kappa$ | **0.875** | **0.825** | **0.700** | **2.80** | **2.55** | **2.20** | **4.3** | **4.1** | **3.5** |
|  |  | 0.0047 | 0.0075 | 0.0103 | 0.0091 | 0.0187 | 0.0373 | 0.0081 | 0.0053 | 0.0244 |
|  | $\chi^2/n$ |  |  |  |  | **2.50** | **2.15** |  |  | **3.4** |

**Table 4.** $D_{eff}$ (bold font) found in the minima of $\chi^2(D_{eff},H_{eff}) / n$ and $\kappa(D_{eff},H_{eff})$ for 1.2 mbar H₂O, for 1.2 mbar C₃H₈ and for 1.2 mbar C₄H₈O for drift-time windows DW-1, DW-2 and DW-3. Missing entries in the row denoted with $\chi^2 / n$ indicate identical values for $D_{eff}$ for the minima of $\chi^2(D_{eff},H_{eff}) / n$ and $\kappa(D_{eff},H_{eff})$. The numbers in normal font are the values of $\kappa(D_{eff},H_{eff})$ in the corresponding minimum. $D_{eff}$ and $H_{eff}$ are jointly varied according to $H_{eff} = F \cdot D_{eff}$ with the ratio $F = H_{eff} / D_{eff}$ obtained by integration of $\eta(r,h|t_c,\Delta t)$. Simulations using GEANT4-DNA were carried out with alpha particles with mean kinetic energy of 3.5 MeV (to simulate an ²⁴¹Am-source sealed with a 10 μm mylar foil) for beam profiles P-1, P-2 +P-2, P-2 −, P-3 and with alpha particles of 3.25 MeV and 3.75 MeV kinetic energy for beams with profile P-2.

Typically, the values of $\kappa(D_{eff},H_{eff})$ are well below 0.02, i.e. the maximum absolute difference between the measured and simulated complementary cumulative ICSDs is less than 2%, indicating a good agreement between measurement and simulation. Only in a few cases this value is larger, but never exceeds 0.06, which still can be regarded a reasonably good agreement. The minima of $\chi^2(D_{eff},H_{eff}) / n$ and $\kappa(D_{eff},H_{eff})$ coincide in 60% of the comparisons, in the other cases the difference in $D_{eff}$ corresponds to the increment in the discrete variation of $D_{eff}$ in the simulation.

Comparison of the minima of $\chi^2(D_{eff},H_{eff}) / n$ and $\kappa(D_{eff},H_{eff})$ shown in Table 4 for the different beam profiles and alpha particles from a sealed ²⁴¹Am-source shows shifts in the cylinder diameter $D_{eff}$ between 2.5% − 5% with respect to the reference profile P-2. The minima of $\chi^2(D_{eff},H_{eff}) / n$ and $\kappa(D_{eff},H_{eff})$ for the beam profile P-2 and alpha particles of 3.25 MeV and 3.75 MeV kinetic energy show shifts in $D_{eff}$ of about 5% with respect to alpha particles from a sealed ²⁴¹Am-source. Therefore, assuming a total uncertainty in the determination of $D_{eff}$ of about ±7% seems justified.

Systematic differences were found in the comparison of ICSDs measured in the three target gases for the three drift-time windows and those obtained by superposition of the track structure simulated in liquid H₂O using GEANT4-DNA, where the corresponding simulated spatial distributions $\eta(r,h|t_c,\Delta t)$ were scaled in terms of liquid H₂O (see Fig. 8). These differences are due to uncertainties in the scaling factors, which depend on $(\rho\lambda_{ion})$-ratios for the different target gases, as well as imperfections in modelling the spatial distribution $\eta(r,h|t_c,\Delta t)$. To investigate the effect of the $(\rho\lambda_{ion})$-ratio on the simulated ICSDs, simulations were carried out varying the ($\rho\lambda_{ion}$)-ratio for a given spatial distribution $\eta(r,h|t_c,\Delta t)$. The results were compared to the corresponding measured ICSDs.

|   |   | H$_2$O | | | C$_3$H$_8$ | | | C$_4$H$_8$O | | |
|---|---|---|---|---|---|---|---|---|---|---|
|   |   | DW-1 | DW-2 | DW-3 | DW-1 | DW-2 | DW-3 | DW-1 | DW-2 | DW-3 |
| ($\rho\lambda_{ion}$) ratio | $\kappa$ | **0.94** | **0.98** | **0.98** | **1.250** | **1.275** | **1.275** | **1.08** | **1.14** | **1.12** |
|   |   | 0.0060 | 0.0063 | 0.0053 | 0.0138 | 0.0115 | 0.0100 | 0.0217 | 0.0229 | 0.0327 |
|   | $\chi^2/n$ |   |   |   | **1.275** |   |   | **1.10** |   | **1.14** |
|   |   | 31.9 | 34.0 | 42.2 | 42.6 | 46.7 | 52.3 | 70.9 | 112.7 | 237.4 |

**Table 5.** ($\rho\lambda_{ion}$)-ratios (bold font) found at the minima of $\chi^2/n$ and $\kappa$ for the comparison between the measured ICSDs and those simulated with GEANT4-DNA using the spatial distributions of the extraction efficiency $\eta(r,h|t_c,\Delta t)$ for 1.2 mbar H$_2$O, for 1.2 mbar C$_3$H$_8$ and for 1.2 mbar C$_4$H$_8$O with drift-time windows DW-1, DW-2 and DW-3. Missing entries in the row denoted with $\chi^2/n$ indicate identical values for the ($\rho\lambda_{ion}$)-ratios for the minima of $\chi^2/n$ and $\kappa$. The numbers printed in normal font are the values of $\kappa$ and $\chi^2/n$ at the corresponding minimum.

Table 5 shows the ($\rho\lambda_{ion}$)-ratios (bold numbers) at the minima of $\chi^2/n$ and $\kappa$ for the comparison between the measured ICSDs and those simulated with GEANT4-DNA using the spatial distributions of the extraction efficiency $\eta(r,h|t_c,\Delta t)$ for 1.2 mbar H$_2$O, for 1.2 mbar C$_3$H$_8$ and for 1.2 mbar C$_4$H$_8$O and drift-time windows DW-1, DW-2 and DW-3. As in Table 4, missing entries in the row denoted with $\chi^2/n$ indicate identical values for the ($\rho\lambda_{ion}$)-ratios for the minima of $\chi^2/n$ and $\kappa$. The numbers printed in normal font are the values of $\kappa$ and $\chi^2/n$ at the corresponding minimum. The corresponding ICSDs are found in the supplement (Fig. S7). The values for $\chi^2/n$ are also listed to allow for comparison with table 2.

The ($\rho\lambda_{ion}$)-ratios at the minima of $\chi^2/n$ and $\kappa$ deviate from those used in the previous scaling: for ($\rho\lambda_{ion}$)$_{H2O,liquid}$ / ($\rho\lambda_{ion}$)$_{H2O,vapour}$ an average value of ≈ 0.967 is found instead of 1.05; for ($\rho\lambda_{ion}$)$_{H2O}$ / ($\rho\lambda_{ion}$)$_{C3H8}$ optimum agreement occurs at a value of ≈ 1.275 rather than of 1.45 and for ($\rho\lambda_{ion}$)$_{H2O}$ / ($\rho\lambda_{ion}$)$_{C4H8O}$ the minima are located at an average value of ≈ 1.11 instead of 1.06. The values found for the ($\rho\lambda_{ion}$)-ratios at the minima of $\chi^2/n$ and $\kappa$ are smaller for H$_2$O and C$_3$H$_8$, deviating by ≈ 10% and ≈ 14%, respectively, and are larger by ≈ 5% for C$_4$H$_8$O. Notably, the value of ($\rho\lambda_{ion}$)$_{H2O}$ / ($\rho\lambda_{ion}$)$_{C3H8}$ at the minima of $\chi^2/n$ and $\kappa$ corresponds to the value of ≈ 1.25 stated in [24].

Optimisation of the ($\rho\lambda_{ion}$)-ratios led to a substantial reduction in $\chi^2/n$ and $\kappa$ compared to the values given in table 2. Optimisation of the ($\rho\lambda_{ion}$)-ratios to obtain minima of $\chi^2/n$ and $\kappa$ compensates for the deviation between measurement and simulation, which result from a combination of effects from three different sources, (i) uncertainty of the ($\rho\lambda_{ion}$)-ratios, (ii) uncertainties due to imperfections in the modelling of the spatial distribution $\eta(r,h|t_c,\Delta t)$ and (iii) uncertainty to which extent the scaling procedure is valid including the secondary electrons in the penumbra. The contributions from these effects cannot be separated, and hence, in some cases they may add up or cancel out each other to some extent. The compensation of these deviations must therefore be regarded with care. However, the difference between the optimised ($\rho\lambda_{ion}$)-ratios and those used in the previous scaling may indicate the possible order of magnitude for the uncertainty of the combined effects.

The effect of the optimised ($\rho\lambda_{ion}$)-ratios on the scaled spatial distributions of the extraction efficiency for the three target gases and the three drift-time windows is shown in Fig. S3 of the supplement.

## 4. Summary and conclusion

For the first time, a dedicated investigation of the target size of a nanodosimeter device has been carried out, with particular focus on the equivalent target size in terms of liquid H$_2$O. As detailed in this work, a method has been developed to characterise the target volume in terms of liquid H$_2$O, not only with respect to the spatial distribution of the extraction efficiency, but also with respect to simplified geometries consisting of a cylindrical and spherical shape, as often used in approaches to model radiation effects to DNA.

Therefore, measurements were carried out using three different target gases, H$_2$O, C$_3$H$_8$ and C$_4$H$_8$O with alpha particles from an $^{241}$Am source. For each of the three target gases, three different drift-time windows were applied leading to three different target sizes.

To determine the degree of agreement between measured and simulated ICSDs, two metrices were applied, the reduced $\chi^2$, i.e. $\chi^2/n$, and the maximum absolute difference $\kappa$ between the two complementary cumulative ICSDs. While the range of the numerical values of $\kappa$ is limited to $0 \le \kappa \lesssim 1$, the range of numerical values for $\chi^2/n$ is not limited but the most likely values for comparison of random samples from the same statistical distribution are in the vicinity of 1. However, considering only the statistical uncertainties of $P_\nu^{EXP}$ and $P_\nu^{MC}$ in the determination of $\chi^2/n$ leads to much larger numerical values for $\chi^2/n$. As the systematic uncertainty contributions are difficult to assess, $\kappa$ may better serve as a quantitative measure of the deviation between the two complementary cumulative ICSDs.

For the three target gases and three drift-time windows, the dimensions of the simulated spatial distribution of the extraction efficiency of the ionised target gas molecules have been scaled to the dimensions of liquid H$_2$O using the scaling procedure described in [24], though strictly speaking, this scaling procedure is only valid for the ionisations due to the primary alpha particles. The measured and simulated complementary cumulative ICSDs agree within $\kappa < 8\%$ for H$_2$O and C$_4$H$_8$O and within $\kappa < 14\%$ for C$_3$H$_8$, likewise do the mean ionisation cluster sizes. This suggests, that at least for the considered irradiation geometry with the primary beam passing the target volume centrally and the considered radiation quality, the dimensional scaling of the spatial distribution of the extraction efficiency is appropriate to estimate to some extent the target volume in liquid H$_2$O. However, the scaling of the spatial distribution of the extraction efficiency with respect to the $(\rho\lambda_{ion})$-ratios is susceptible to several sources of uncertainties, such as (i) the uncertainty of the $(\rho\lambda_{ion})$-ratios, (ii) uncertainties in the modelling of the spatial distribution $\eta(r,h|t_c,\Delta t)$ and (iii) the uncertainty to which extent the scaling procedure is also valid for the secondary electrons in the penumbra. Optimisation of the $(\rho\lambda_{ion})$-ratios substantially improves the agreement of the complementary cumulative measured and simulated ICSDs for all target gases and drift-time windows. This may indicate that the possible order of magnitude for the uncertainty of the combined effects in the present investigation is as much as 5% for C$_4$H$_8$O, 10% for H$_2$O and 15% for C$_3$H$_8$. Notably, the value found for $(\rho\lambda_{ion})_{H2O}$ / $(\rho\lambda_{ion})_{C3H8}$ at the minima of $\chi^2$ / $n$ and $\kappa$ corresponds to the value of $\approx 1.25$ stated in [24], indicating that the revision of the cross sections for C$_3$H$_8$ in [25] may have led to an overestimation of the ionisation cross sections for alpha particles.

For simulations with PTra using the spatial distribution of the extraction efficiency for C$_3$H$_8$, both $\chi^2$ / $n$ and $\kappa$ are smaller by a factor of about 2.5 than those obtained for simulations with GEANT4-DNA using the spatial distribution of the extraction efficiency scaled to liquid H$_2$O. Similarly, for the mean ionisation cluster sizes $M_1$, the values obtained from simulations with PTra agree significantly better with measured $M_1$ values than those obtained from simulations with GEANT4-DNA. PTra and GEANT4-DNA, however, use different cross sections and transport models for ionising particles and the track structure simulated with GEANT4-DNA is superimposed with a target volume of dimensions scaled to liquid H$_2$O using a ratio $(\rho\lambda_{ion})_{H2O}$ / $(\rho\lambda_{ion})_{C3H8}$ = 1.45. All together this doesn't allow to exactly pinpoint the origin of this difference.

To determine the equivalent target size in terms of liquid H$_2$O for cylindrical and spherical targets, track structures simulated with GEANT4-DNA in a slab of liquid H$_2$O were superimposed with target volumes of respective shape, and the resulting ICSDs were compared to ICSDs measured in the three target gases for the three drift-time windows. For a quantitative determination of the degree of agreement the two metrics $\chi^2(D_{eff},H_{eff})$ / $n$ and $\kappa(D_{eff},H_{eff})$ were applied. The values of both $\chi^2(D_{eff},H_{eff})$ / $n$ and $\kappa(D_{eff},H_{eff})$ varied only slightly along the bottom of the valley. Due to the shape of $\chi^2(D_{eff},H_{eff})$ / $n$ and $\kappa(D_{eff},H_{eff})$, a unique determination of $D_{eff}$ and $H_{eff}$ was impossible. Therefore, additional information is required for a unique determination, which is obtained by integration of the spatial distribution of the extraction efficiency yielding the ratio $H_{eff}$ / $D_{eff}$. Applying this ratio in the comparison of measured and simulated ICSDs led to unique and reasonable results for $D_{eff}$ and $H_{eff}$ which are consistent with the shape of both $\chi^2(D_{eff},H_{eff})$ / $n$ and $\kappa(D_{eff},H_{eff})$. Measured and simulated ICSDs as well as the mean ionisation cluster sizes agreed very well. The deviation of the cumulative complementary ICSDs was mostly below 1.5% and only about 4.0% in the worst case. Measurement and simulation were almost coincident over the whole range of relevant cluster sizes. The mean ionisation cluster sizes $M_1$ agreed within 2% for all combinations of target gas and drift-time window. Thus, by applying the metrics $\chi^2(D_{eff},H_{eff})$ / $n$ and $\kappa(D_{eff},H_{eff})$ and the ratio $H_{eff}$ / $D_{eff}$ both $H_{eff}$ and $D_{eff}$ can be uniquely determined.

For the shortest drift-time window with an almost spherical shape, the metrices $\chi^2(D_{eff})$ / $n$ and $\kappa(D_{eff})$ can be applied to determine the diameter $D_{eff}$ of a spherical target volume. Again, measured and simulated ICSDs and the mean ionisation cluster sizes agreed very well. The deviation of the cumulative complementary ICSDs is mostly below 1% and only about 2.5% in the worst case. Measurement and simulation are almost coincident over the whole range of relevant cluster sizes. The mean ionisation cluster sizes $M_1$ also agreed on the sub-percent level for all target gases.

Thus, the target volume of the nanodosimeter can be well approximated by a cylindrical target volume, and, for short drift-time windows, by a spherical target volume.

Investigation of the effect of beam profiles and the variation of the cross section for ionisation lead to an estimation of the total uncertainty of approximately ±7% in the determination of $D_{eff}$.

Each of the three target gases covers another range of cylindrical target volume, as listed in table 3. Adjusting the length of the drift-time window allows each volume within the respective range to be realised. For the shortest drift-time window, each of the three target gases realises a spherical target volume of a different diameter $D_{eff}$. Variation of the target gas density enables an extension of the respective ranges of the dimensions in terms of liquid H$_2$O of the target volume for each target gas: increasing the density will lead to an increase towards larger dimensions and decreasing the density will lead to an decrease towards smaller dimensions.

ICSDs measured with a nanodosimeter device are therefore suited as benchmark data for approaches that model radiation induced damage to DNA in nanometric volumes using liquid H$_2$O as a target material, even in simple geometries like cylinders or spheres, provided that the nanodosimeter's target volume is characterised accordingly. By proper selection of the target gas and density as well as the length of the drift-time window, a wide range of dimensions in terms of liquid H$_2$O of the target volume can be realised according to particular requirements in the modelling approaches.

Strictly speaking, the findings presented in this paper are only valid for one single irradiation geometry, i.e. primary particles passing the target volume centrally, and one single radiation quality, i.e. alpha particles of 3.5 MeV. An investigation to extend this study to other radiation qualities and irradiation geometries is in progress with particular focus on radiation geometries where ionisation events in the target volume are exclusively due to secondary electrons in the penumbra of the primary particle's track.

**Acknowledgement**
The authors gratefully acknowledge the developers of the nanodosimeter from the Weizmann Institute of Science, Rehovot, Israel, for transferring the device as described in [20] to PTB for further use. The authors also express their gratitude to B. Lambertsen and A. Pausewang for their invaluable contributions to preparation and carrying out the measurements and their assistance in data processing. H. Nettelbeck is acknowledged for proof-reading the manuscript.

**Supplement**

## S1. Validation of the spatial distribution of the extraction efficiency

Validation of the obtained spatial distribution was achieved by comparison of the simulated image of $\eta(r,h|t_c,\Delta t)$ for the corresponding drift time windows in the plane of the PSD surface. Essentially, the simulation of the image of $\eta(r,h|t_c,\Delta t)$ consists of Monte-Carlo integrations of $p(r,z,t_d)$ along the primary particle's trajectory, which ends on a specified virtual pixel on the PSD surface. To approximate the experimental conditions as close as possible, the simulation of the image on the PSD includes the influences of (i) the geometrical setup, i.e. position and size of ion source and virtual detecting pixel, (ii) the position resolution of the two-dimensional PSD and (iii) the radial distribution of ionisations due to secondary electrons in the penumbra of the $^{241}$Am alpha particle tracks.

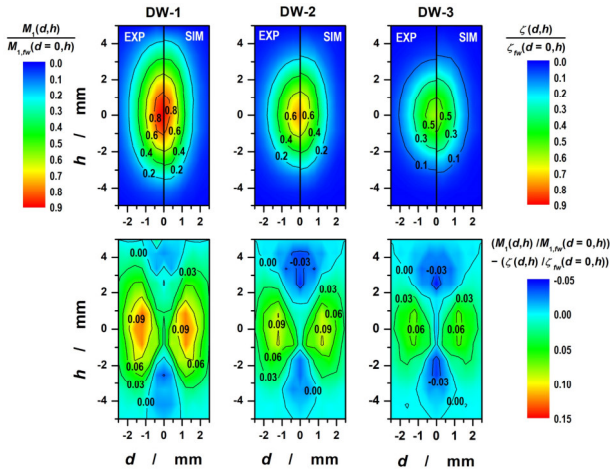

**Figure S1.** Contour plots of the targets for the three drift-time windows DW-1 (left), DW-2 (middle) and DW-3 (right) for 1.2 mbar $C_3H_8$. The plots in the upper row show for $d < 0$ the distribution of the mean ionisation cluster size $M_1(d,h)$ measured with $^{241}$Am alpha particles ("EXP") and for $d > 0$ the simulated image $\zeta(d,h)$ of the simulated extraction efficiency $\eta(r,h|t_c,\Delta t)$ ("SIM"). Each of these distributions has been normalised to the value obtained on the central axis $(d = 0,h)$ for the complete drift time distribution (full window, "*fw*"), i.e. $M_1(d,h) / M_{1,fw}(d = 0,h)$ and $\zeta(d,h) / \zeta_{fw}(d = 0,h)$, respectively. The plots in the bottom row show the difference between the measured and simulated distributions, $M_1(d,h) / M_{1,fw}(d = 0,h) - \zeta(d,h) / \zeta_{fw}(d = 0,h)$.

This is illustrated in Fig. S1 where the upper row of panels shows exemplarily the measured spatial distributions of the mean ionisation cluster size $M_1(d,h)$ of the target volume produced by alpha particles from an $^{241}$Am-source in 1.2 mbar $C_3H_8$ together with the corresponding simulated image $\zeta(d,h)$ of the spatial distribution of the extraction efficiency $\eta(r,h)$ for the three drift-time windows DW-1 (left), DW-2 (middle) and DW-3 (right). $M_1(d,h)$ is normalised to $M_{1,fw}(d = 0,h)$ obtained for the complete (full window, "*fw*") drift time distribution on the central axis at $(d = 0,h)$, and $\zeta(d,h)$ is normalised to the respective quantity $\zeta_{fw}(d = 0,h)$ of the simulated image of the extraction efficiency. For all three drift-time windows, the measured and simulated distributions are in good qualitative agreement and the maxima of the normalised $M_1(d,h)$ and $\zeta(d,h)$ are almost identical.

However, the measured distributions are somewhat wider in $d$ than the simulated ones, while the opposite is observed with respect to the extension in $h$: here the simulated distributions extend a little more in $h$ than the measured ones. This is further illustrated by the contour plots in the bottom row of Fig. S1 that show the differences between the measured and simulated distributions. The maximum deviations can be seen in the border regions of the spatial distributions for $d \approx 1.5$ mm around $h \approx 0$ mm and on the central axis for $h \approx -3$ mm and $h \approx 3$ mm.

## S2. Scaled spatial distributions of the extraction efficiency

Fig. S2 shows contour plots of central cross-sections of the simulated extraction efficiency $\eta(r,h|t_c,\Delta t)$ for the three drift-time windows and the three target gases. The spatial dimensions are scaled in terms of liquid $H_2O$ using the scaling factors $0.93 \cdot 10^{-6}$ for $H_2O$ vapour, $3.15 \cdot 10^{-6}$ for $C_3H_8$ and $3.71 \cdot 10^{-6}$ for $C_4H_8O$.

From the comparison of the plots of DW-1 for 1.2 mbar $C_3H_8$ and DW-3 for 1.2 mbar $C_3H_8O$ ICSDs with larger

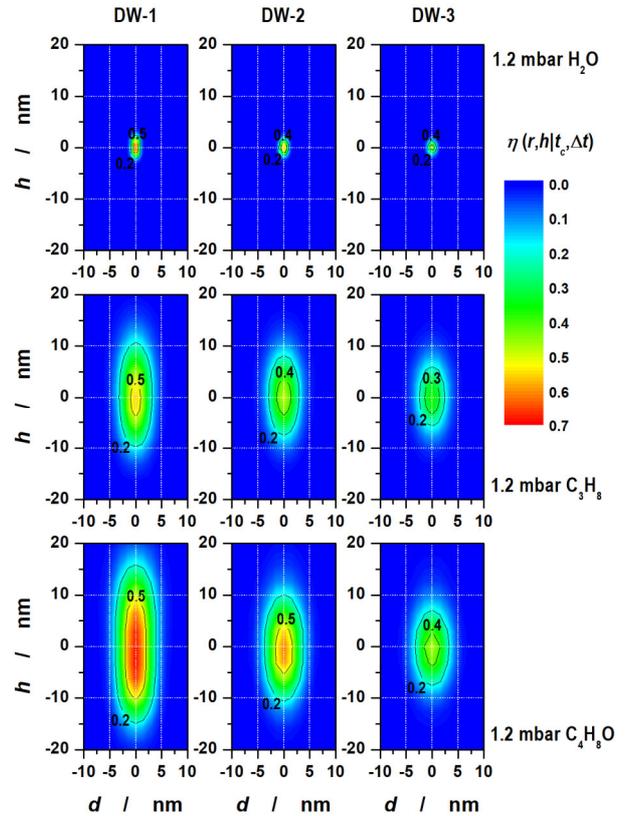

**Figure S2.** Contour plots of the central cross-sections of the simulated extraction efficiency $\eta(r,h|t_c,\Delta t)$ for the drift-time windows DW-1 (left), DW-2 (middle), and DW-3 (right) for 1.2 mbar $H_2O$ (top), 1.2 mbar $C_3H_8$ (middle) and 1.2 mbar $C_4H_8O$ (bottom). The spatial dimensions are scaled in terms of liquid $H_2O$ using the scaling factors $0.93 \cdot 10^{-6}$ for $H_2O$ vapour, $3.15 \cdot 10^{-6}$ for $C_3H_8$ and $3.71 \cdot 10^{-6}$ for $C_4H_8O$.

$M_1$ values are expected for DW-1 and 1.2 mbar $C_3H_8$ than for DW-3 and 1.2 mbar $C_3H_8O$ due to the comparable diameter but larger extraction efficiency in the centre of DW-1 for 1.2 mbar $C_3H_8$ compared to DW-3 for 1.2 mbar $C_3H_8O$. In fact, this is the case for the simulated but not for the measured ICSDs, as can be seen in Fig. 8. Applying the $(\rho\lambda_{ion})$-ratios from table 5 in the scaling of the spatial dimensions of the extraction efficiency leads to the contour plots shown in Fig. S3. Here, the scaling factors are $0.86\cdot10^{-6}$ for $H_2O$ vapour, $2.77\cdot10^{-6}$ for $C_3H_8$ and $3.89\cdot10^{-6}$ for $C_4H_8O$, leading to a downsized target volume for $H_2O$ vapour and $C_3H_8$ and an enlarged target size for $C_4H_8O$. This is reflected in the corresponding ICSDs (see Fig. S7).

## S3. Simulations for liquid H₂O with GEANT4-DNA

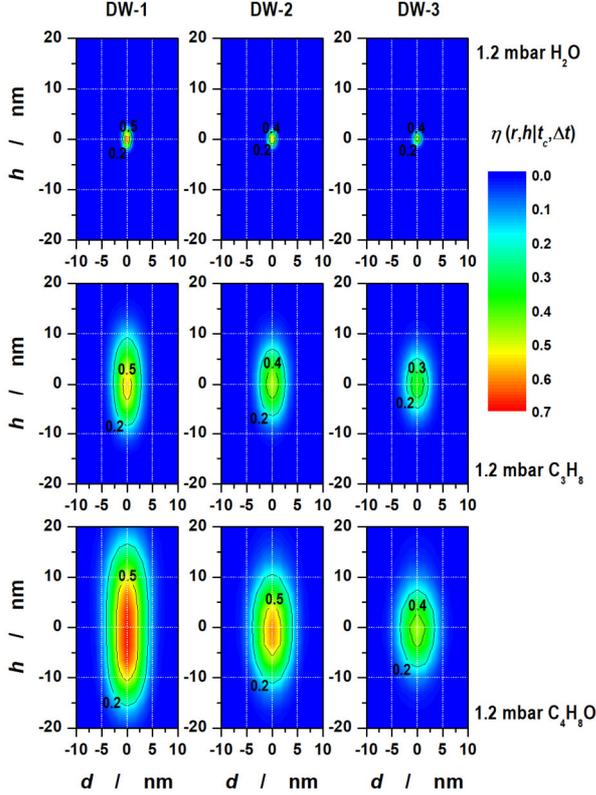

**Figure S3.** Contour plots of the central cross-sections of the simulated extraction efficiency $\eta(r,h|t_c,\Delta t)$ for the drift-time windows DW-1 (left), DW-2 (middle), and DW-3 (right) for 1.2 mbar $H_2O$ (top), 1.2 mbar $C_3H_8$ (middle) and 1.2 mbar $C_4H_8O$ (bottom) applying the $(\rho\lambda_{ion})$-ratios from table 5. The spatial dimensions are scaled in terms of liquid $H_2O$ using the scaling factors $0.86\cdot10^{-6}$ for $H_2O$ vapour, $2.77\cdot10^{-6}$ for $C_3H_8$ and $3.89\cdot10^{-6}$ for $C_4H_8O$.

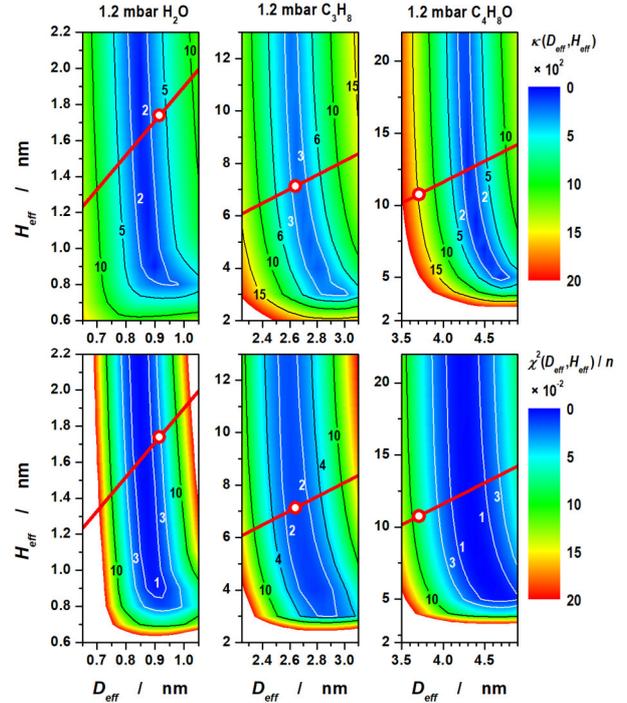

**Figure S4.** Contour plots of $\chi^2(D_{eff},H_{eff})/n$ and $\kappa(D_{eff},H_{eff})$ for the comparison of ICSDs measured in 1.2 mbar $H_2O$ (left), in 1.2 mbar $C_3H_8$ (middle) and in 1.2 mbar $C_4H_8O$ (right) for the drift-time window DW-2 and the corresponding ICSDs simulated with GEANT4-DNA using an ideal target cylinder of liquid $H_2O$ with dimensions $D_{eff}$ and $H_{eff}$. The red circle with the white centre marks $D_{eff}$ and $H_{eff}$, scaled in terms of liquid $H_2O$, obtained by integration of $\eta(r,h|t_c,\Delta t)$ for the drift-time window DW-2. Along the red line, the ratio of $D_{eff}/H_{eff}$ is the same as it is obtained by integration of $\eta(r,h|t_c,\Delta t)$.

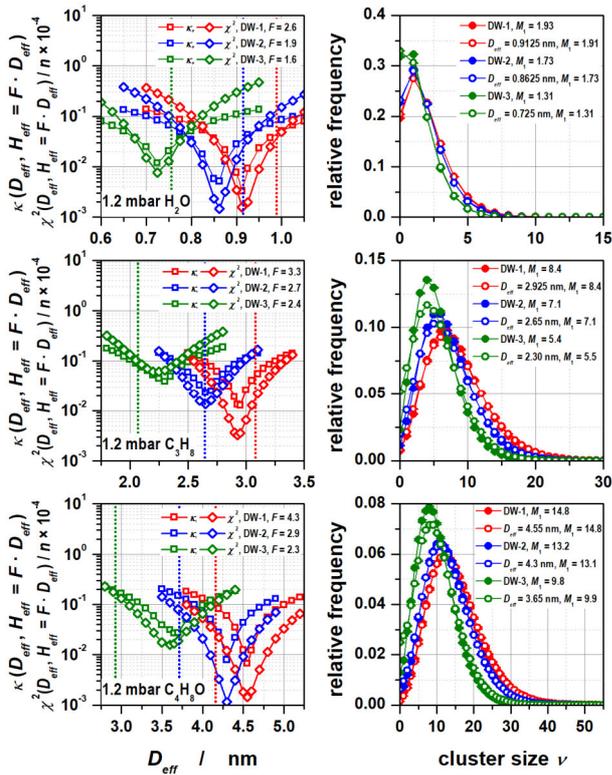

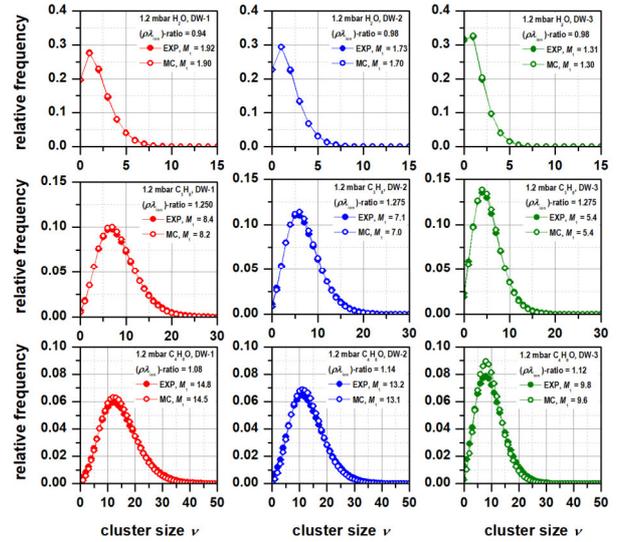

**Figure S5.** $\chi^2(D_{eff},H_{eff}) / n$ and $\kappa(D_{eff},H_{eff})$ for 1.2 mbar H$_2$O (top row), for 1.2 mbar C$_3$H$_8$ (middle row) and for 1.2 mbar C$_4$H$_8$O (bottom row) for the drift-time windows DW-1, DW-2 and DW-3. $D_{eff}$ and $H_{eff}$ are jointly varied according to $H_{eff} = F \cdot D_{eff}$ with the ratio $F = H_{eff} / D_{eff}$ obtained by integration of $\eta(r,h|t_c,\Delta t)$. The vertical dotted lines indicate the values of $D_{eff}$ obtained by integration of $\eta(r,h|t_c,\Delta t)$ and scaled to liquid H$_2$O. Right: ICSDs measured in the respective target gas for the drift-time windows DW-1, DW-2 and DW-3 (full circles) and the corresponding ICSDs simulated with GEANT4-DNA using an ideal cylinder of liquid H$_2$O with the values found for $D_{eff}$ and $H_{eff}$ at the minimum of $\kappa(D_{eff},H_{eff})$ for the respective drift-time window (open circles).

**Figure S7.** ICSDs measured ("EXP") in 1.2 mbar H$_2$O (top row), in 1.2 mbar C$_3$H$_8$ (middle row) and in 1.2 mbar C$_4$H$_8$O (bottom row) for the three drift-time windows DW-1 (red, left column), DW-2 (blue, middle column) and DW-3 (green, right column) and simulated ICSDs ("MC"). The latter were obtained from superposition of the track structure simulated in liquid H$_2$O with the corresponding simulated spatial distributions $\eta(r,h|t_c,\Delta t)$ scaled in terms of liquid H$_2$O by applying the $(\rho\lambda_{ion})$-ratios listed in table 5.

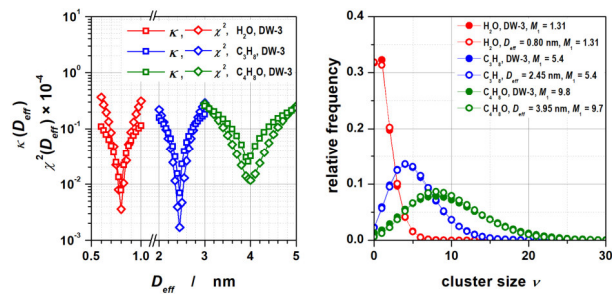

**Figure S6.** Left: $\chi^2(D_{eff},H_{eff}) / n$ and $\kappa(D_{eff})$ for 1.2 mbar H$_2$O (red), for 1.2 mbar C$_3$H$_8$ (blue) and for 1.2 mbar C$_4$H$_8$O (green) for the drift-time window DW-3. Right: ICSDs measured in the respective target gas for the drift-time window DW-3 (full circles) and the corresponding ICSDs simulated with GEANT4-DNA using an ideal sphere of liquid H$_2$O with the values found for $D_{eff}$ at the minimum of $\kappa(D_{eff})$ (open circles).

## S4. Effect of the statistical uncertainties of $P_\nu^{EXP}$ and $P_\nu^{MC}$ on the numerical values of $\chi^2 / n$

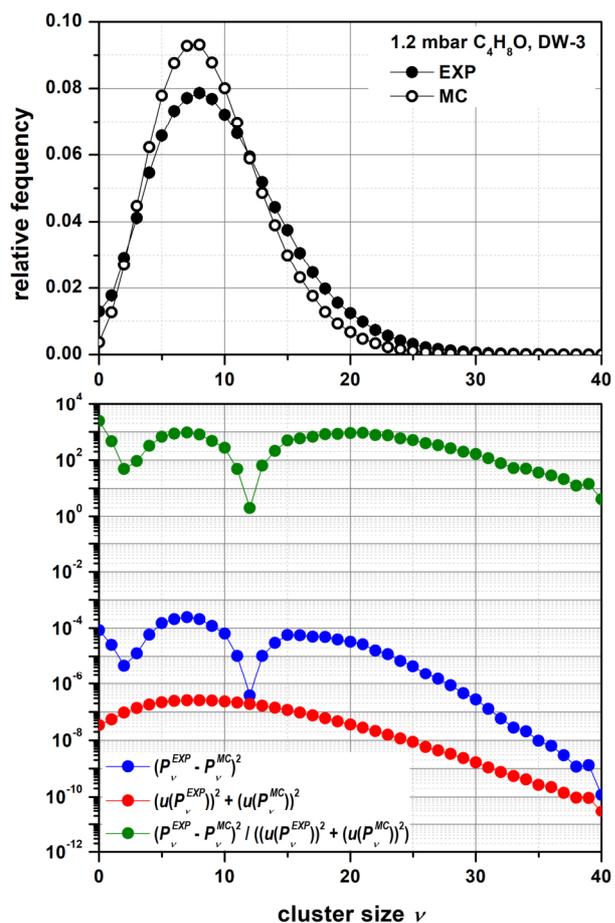

**Figure S8.** Effect of including the squares $(u(P_\nu^{EXP}))^2$ and $(u(P_\nu^{MC}))^2$ of the statistical uncertainties of $P_\nu^{EXP}$ and $P_\nu^{MC}$ on the numerical values of $\chi^2 / n$, exemplarily shown for the comparison of measured (full circles) and simulated (open circles) ICSDs for 1.2 mbar $C_4H_8O$ and DW-3. The upper plot shows the ICSDs, the lower plot shows the numerical values of $(P_\nu^{EXP} - P_\nu^{MC})^2$ (blue), $(u(P_\nu^{EXP}))^2 + (u(P_\nu^{MC}))^2$ (red) and $(P_\nu^{EXP} - P_\nu^{MC})^2 / ((u(P_\nu^{EXP}))^2 + (u(P_\nu^{MC}))^2)$ (green).